\documentstyle[sprocl]{article}
 
\newcommand{\nc}{\newcommand}
\nc{\rem}[1]{{\bf [#1]}}
\nc{\eqn}[1]{Eq.~\ref{eq:#1}}
\nc{\eqns}[2]{Eqs.~\ref{eq:#1} and \ref{eq:#2}}
\nc{\fig}[1]{Fig.~\ref{fig:#1}}
\nc{\figs}[2]{Figs.~\ref{fig:#1} and \ref{fig:#2}}
\nc{\be}{\begin{equation}}
\nc{\ee}{\end{equation}}
\nc{\ba}{\begin{array}}
\nc{\ea}{\end{array}}
\nc{\bea}{\begin{eqnarray}}
\nc{\eea}{\end{eqnarray}}
\nc{\del}{\partial}
\nc{\eqconst}{\, \mbox{(const.)}\, }
\nc{\eqand}{\quad\mbox{and}\quad}
\nc{\eqRe}{\mbox{Re}\, }
\nc{\eqIm}{\mbox{Im}\, }
\nc{\alt}{\buildrel < \over {_\sim}}
\nc{\agt}{\buildrel > \over {_\sim}}
\nc{\st}{\scriptstyle}
\nc{\sst}{\scriptscriptstyle}
\nc{\mco}{\multicolumn}
\nc{\vep}{\varepsilon}
\nc{\lr}{\leftrightarrow}
\nc{\ra}{\rightarrow}
\nc{\vp}{{\bf p}}
\nc{\al}{\alpha}
\nc{\as}{\alpha_s}
\nc{\ab}{\bar{\alpha}}
\nc{\semi}{;\ }
\nc{\spa}[3]{\left\langle#1\,#3\right\rangle}
\nc{\spb}[3]{\left[#1\,#3\right]}
\nc{\LP}{\left(}
\nc{\RP}{\right)}
\nc{\LB}{\left[}
\nc{\RB}{\right]}
\nc{\Tr}{\mathop{\rm Tr}\nolimits}
\nc{\tr}{\mathop{\rm tr}\nolimits}
\nc{\e}{\epsilon}
\nc{\cg}{c_\Gamma}
\nc{\hf}{\textstyle{1\over2}}
\nc{\Li}{\mathop{\rm Li}\nolimits}
\nc{\Ls}{\mathop{\rm Ls}\nolimits}
\nc{\Ll}{\mathop{\rm L}\nolimits}
\nc{\gluino}{{\tilde g}}
\nc{\tg}{{\tilde g}}
\nc{\tq}{{\tilde q}}
\nc{\qb}{{\bar q}}
\nc{\Qb}{{\bar Q}}
\nc{\SUSY}{{\rm SUSY}}
\nc{\susy}{{\rm SUSY}}
\nc{\tree}{{\rm tree}}
\nc{\oneloop}{{\rm 1-loop}}
\nc{\A}{{\cal A}}
\nc{\Atree}{A^{\rm tree}}
\nc{\Atreestar}{A^{\rm tree\,*}}
\nc{\Aloop}{A^{\rm 1-loop}}
\nc{\Aloopstar}{A^{\rm 1-loop\,*}}
\nc{\pol}{\varepsilon}
\nc{\si}{\sigma}
\nc{\ns}{n_{\mskip-2mu s}}
\nc{\nf}{n_{\mskip-2mu f}}
\nc{\ib}{{\bar\imath}}
\nc{\jb}{{\bar\jmath}}
\nc{\treemhv}{{\rm tree\ MHV}}
\nc{\loopmhv}{{\rm 1-loop\ MHV}}
\nc{\dlips}{d{\rm LIPS}}
\nc{\lsl}{\not{\hbox{\kern-2.3pt $\ell$}}}
\nc{\ksl}{\not{\hbox{\kern-2.3pt $k$}}}
\nc{\esl}{\not{\hbox{\kern-2.3pt $\pol$}}}
\nc{\Psl}{\not{\hbox{\kern-2.3pt $P$}}}
\nc{\Dsl}{\not{\hbox{\kern-2.3pt $D$}}}
\nc{\Slash}[1]{\slash\hskip -0.17 cm #1}
\nc{\tn}[2]{t^{[#1]}_{#2}}
\nc{\Split}{\mathop{\rm Split}\nolimits}
\nc{\Soft}{\mathop{\rm Soft}\nolimits}
\input epsf
\nc{\figdir}{} 
\nc{\figmac}[5]{\begin{figure}
\centerline{\parbox[t]{#1in}{\epsfbox{\figdir #2.ps}}}
\caption[#4]{\label{fig:#3} #5}\end{figure}}
\def\epsfsize#1#2{\ifdim#1>\hsize\hsize\else#1\fi}
\begin{document}

\pagestyle{empty}
\vspace{-1.1in}
\rightline{SLAC--PUB--7106}
\rightline{hep-ph/9601359}
\rightline{January, 1996}

\vspace{1.0cm}
\centerline{\bf CALCULATING SCATTERING AMPLITUDES 
                      EFFICIENTLY${}^{\star}$
}

\vspace*{9.0ex}
\centerline{\rm LANCE DIXON}
\vspace*{1.5ex}
\centerline{\it Stanford Linear Accelerator Center}
\vspace*{1.0ex}
\centerline{\it Stanford University, Stanford, CA 94309}
\vspace*{4.5ex}
\vskip3.0truecm

\centerline{\small ABSTRACT}
\vspace*{1.0ex}
\abstracts{ 
We review techniques for more efficient computation of perturbative
scattering amplitudes in gauge theory, in particular
tree and one-loop multi-parton amplitudes in QCD.
We emphasize the advantages of (1) using color and helicity information
to decompose amplitudes into smaller gauge-invariant pieces, and
(2) exploiting the analytic properties of these pieces, 
namely their cuts and poles.  Other useful tools include
recursion relations, special gauges and supersymmetric 
rearrangements. 
}

\vspace*{3.0ex}

\centerline{\sl Invited lectures presented at the 
                Theoretical Advanced Study Institute}
\centerline{\sl in Elementary Particle Physics (TASI 95): 
                QCD and Beyond}
\centerline{\sl Boulder, CO, June 4-30, 1995}

\vfil\vskip .2 cm
\noindent\hrule width 3.6in\hfil\break
${}^{\star}${\small 
Research supported by the US Department of Energy
under grant DE-AC03-76SF00515.}
\eject

\pagestyle{plain}
 
\section{Motivation}

Feynman rules for covariant perturbation theory have been around for 
almost fifty years, and their adaptation to nonabelian gauge theories
has been fully developed for almost twenty-five years.
Surely by now every significant standard model scattering 
process ought to have been calculated to the experimentally-required 
accuracy.
In fact, this is far from the case, especially for QCD, which is the 
focus of this school and of these lectures.
Many QCD cross-sections have been calculated only to leading 
order (LO) in the strong coupling constant $\as$, 
corresponding to the square of the tree-level amplitude.   
Such calculations have very large 
uncertainties --- often a factor of two --- which can only be reduced
to reasonable levels, say 10\% or so, by including higher-order 
corrections in $\as$.   

Currently, no quantities have been computed beyond
next-to-next-to-lead\-ing-order (NNLO) in $\as$, 
and the only quantities known at NNLO are totally inclusive quantities
such as the total cross-section for $e^+e^-$ annihilation into hadrons,
and various sum rules in deep inelastic
scattering.  Many more processes have been calculated at 
next-to-leading-order (NLO), but at present
results are still limited to where the basic process has
four external legs, such as a virtual photon or $Z$ decaying to three
jets, or production of a pair of jets (or a weak boson plus a jet)
in hadronic collisions via $q\qb\to gg$ ($q\qb\to Wg$), etc.

This is not to say that processes with more external legs are not
interesting; they are of much interest, both for testing QCD in different 
settings and as backgrounds to new physics processes.
For example, $\alpha_s$ could be measured at the largest possible
momentum transfers using the ratio of three-jet events to two-jet
events at hadron colliders, if only the three-jet process were known 
at NLO. 
As another example, QCD is a major background to top quark production 
in $p\bar{p}$ collisions.  If both $t$'s decay hadronically 
($t\to Wb\to q\qb^\prime b$), the background is from six jet 
production.  Despite the fact that the QCD process starts 
off at $\as^6$, it completely swamps the top signal.  
If one of the two top quarks decays leptonically 
($t\to Wb\to \bar{\ell}\nu_\ell b$), then QCD production of a 
$W$ plus three or four jets forms the primary background.
This background prevented discovery of the top quark at the Tevatron
in this channel, until the advent of $b$ tagging.\cite{topdiscovery}
Although the NLO corrections to three-jet production are within sight,
we are still far from being able to compute the top quark backrounds 
at NLO accuracy; 
on the other hand, it's good to have long range goals.

These lectures are about amplitudes rather than cross-sections.
The goal of the lectures is to introduce you to efficient 
techniques for computing tree and one-loop amplitudes in QCD,
which serve as the input to LO and NLO cross-section calculations.
(The same techniques can be applied to many non-QCD 
multi-leg processes as well.)
Zoltan Kunszt will then describe in detail how to combine amplitudes 
into cross-sections.\cite{KunsztTASI}

Efficient techniques for computing tree amplitudes have been 
available for several years, and an excellent review 
exists.\cite{ManganoReview}
One-loop calculations are considerably more involved --- they form an
``analytical bottleneck'' to obtaining new NLO results --- 
and benefit from additional techniques.
In principle it is straightforward to compute both tree and loop 
amplitudes by drawing all Feynman diagrams and evaluating them, using 
standard reduction techniques for the loop integrals that are
encountered.  In practice this method becomes extremely inefficient 
and cumbersome as the number of external legs grows, because there are:
\par\noindent
1. {\bf too many diagrams} --- many diagrams are related by gauge 
invariance.
\par\noindent
2. {\bf too many terms in each diagram} --- nonabelian gauge boson 
self-interactions are complicated. 
\par\noindent
3. {\bf too many kinematic variables} --- allowing the construction of
arbitrarily complicated expressions.
\par\noindent
Consequently, intermediate expressions tend to be vastly more 
complicated than the final results, when the latter are
represented in an appropriate way.

In these lectures we will stress the advantages of
(1) using color and helicity information to decompose amplitudes 
into smaller (and simpler) gauge-invariant pieces, and
(2) exploiting the analytic properties of these pieces, 
namely their cuts and poles.
In this way one can tame the size of intermediate expressions as
much as possible on the way to the final answer.   
There are many useful technical steps and tricks along the way,
but I believe the overall organizational philosophy is just as
important.  A number of the techniques can be motivated 
by how calculations are organized in string 
theory.\cite{StringBased,Long}
I will not attempt to describe string theory here, but I will mention 
some places where it provides a useful heuristic guide.
 
The approach advocated here is quite
useful for multi-parton scattering amplitudes.   
For more inclusive processes --- for example the
$e^+e^-\to$~hadrons total cross-section --- where the number of
kinematic variables is smaller, and the real and virtual contributions
are on a more equal footing, the computational issues
are completely different, and the philosophy of splitting the problem
up into many pieces may actually be counterproductive.


\section{Total quantum-number management (TQM)}

The organizational framework mentioned above uses all the
quantum-numbers of the external states (colors and helicity) to
decompose amplitudes into simpler pieces; thus we might dub it
``Total Quantum-number Management''.  TQM suggests that we:
\par\noindent
$\bullet$ Keep track of all possible information about external
particles --- namely, {\it helicity} and {\it color} information.
\par\noindent
$\bullet$ Keep track of quantum {\it phases} by computing the
transition amplitude rather than the cross-section.
\par\noindent
$\bullet$ Use the helicity/color information to decompose the amplitude
into simpler, gauge-invariant pieces, called {\it sub-amplitudes} 
or {\it partial amplitudes}.  
\par\noindent
$\bullet$ In many cases we may also introduce still simpler auxiliary
objects, called {\it primitive amplitudes}, out of which the
partial amplitudes are built.
\par\noindent
$\bullet$ Exploit the ``effective'' {\it supersymmetry} of QCD tree
amplitudes, and use supersymmetry at loop-level to help manage the 
spins of particles propagating around the loop.
\par\noindent
$\bullet$ Square amplitudes to get probabilities, and sum over 
helicities and colors to obtain unpolarized cross-sections, 
only at the very {\it end} of the calculation.  
\par\noindent
Carrying out the last step explicitly would generate a large 
analytic expression; however, at this stage one would
typically make the transition to numerical evaluation, 
in order to combine the virtual and real corrections.
The use of TQM is hardly new, particularly in tree-level
applications\,\cite{ManganoReview} ---
but it becomes especially useful at loop level.

\subsection{Color management}

First we describe the color decomposition of
amplitudes,\cite{TreeColor,LoopColor} and review
some diagrammatic techniques\,\cite{DoubleLine}
for efficiently carrying out the necessary group theory.
The gauge group for QCD is $SU(3)$, but there is no harm in
generalizing it to $SU(N_c)$; indeed this makes some of the group
theory structure more apparent.
Gluons carry an adjoint color index $a=1,2,\ldots,N_c^2-1$, 
while quarks and antiquarks carry an $N_c$ or $\overline{N}_c$ index,
$i,\jb=1,\ldots,N_c$.  
The generators of $SU(N_c)$ in the fundamental representation are
traceless hermitian $N_c\times N_c$ matrices, $(T^a)_i^{~\jb}$.
We normalize them according to $\Tr(T^aT^b) = \delta^{ab}$
in order to avoid a proliferation of $\sqrt{2}$'s in partial 
amplitudes.   (Instead the $\sqrt{2}$'s appear in intermediate steps
such as the color-ordered Feynman rules in \fig{RulesFigure}.)

The color factor for a generic Feynman diagram in QCD contains
a factor of $(T^a)_i^{~\jb}$ for each gluon-quark-quark vertex,
a group theory structure constant $f^{abc}$ --- defined by
$[T^a,T^b]\ =\ i\sqrt{2}\,f^{abc}\,T^c$ --- for each pure gluon 
three-vertex, and contracted pairs of structure constants
$f^{abe}f^{cde}$ for each pure gluon four-vertex. 
The gluon and quark propagators contract many of the indices
together with $\delta_{ab}$, $\delta^{~\jb}_{i}$ factors.
We want to first identify all the different types of color factors
(or ``color structures'') that can appear in a given amplitude,
and then find rules for constructing the kinematic coefficients
of each color structure, which are called sub-amplitudes or partial
amplitudes.

The general color structure of the amplitudes can be exposed if we 
first eliminate the structure constants $f^{abc}$ in favor of the 
$T^a$'s, using
\be \label{eq:struct}
f^{abc}\ =\ -{i\over\sqrt2}\Bigl( 
 \Tr\bigl(T^aT^bT^c\bigr) - \Tr\bigl(T^aT^cT^b\bigr) \Bigr),
\ee
which follows from the definition of the structure constants.
At this stage we have a large number of traces, many sharing 
$T^a$'s with contracted indices, of the form 
$\Tr\bigl(\ldots T^a\ldots\bigr)\,\Tr\bigl(\ldots T^a\ldots\bigr)
\,\ldots\,\Tr\bigl(\ldots)$.
If external quarks are present, then in addition to the traces there
will be some strings of $T^a$'s terminated by fundamental indices,
of the form $(T^{a_1}\ldots T^{a_m})_{i_2}^{~\ib_1}$. 
To reduce the number of traces and strings we ``Fierz rearrange'' 
the contracted $T^a$'s, using
\be \label{eq:colorfierz}
  (T^a)_{i_1}^{~\jb_1} \, (T^a)_{i_2}^{~\jb_2}\ =\ 
  \delta_{i_1}^{~\jb_2} \delta_{i_2}^{~\jb_1}
  - {1\over N_c} \, \delta_{i_1}^{~\jb_1} \delta_{i_2}^{~\jb_2}\,,
\ee
where the sum over $a$ is implicit.  

Equation~\ref{eq:colorfierz} is just the statement that the $SU(N_c)$
generators $T^a$ form the complete set of traceless hermitian 
$N_c\times N_c$ matrices.  The $-1/N_c$ term implements the
tracelessness condition.  (To see this, contract both sides
of \eqn{colorfierz} with $\delta^{~i_1}_{\jb_1}$.) 
It is often convenient to consider also
$U(N_c)\ =\ SU(N_c) \times U(1)$ gauge theory.  The additional $U(1)$ 
generator is proportional to the identity matrix, 
\be \label{eq:photongenerator}
(T^{a_{U(1)}})_{i}^{~\jb} = {1 \over \sqrt{N_c}}\ \delta_{i}^{~\jb}\ ;
\ee
when this is added back the
$U(N_c)$ generators obey \eqn{colorfierz} without the $-1/N_c$ term.   
The auxiliary $U(1)$ gauge field is often called the photon, because
it is colorless 
(it commutes with $SU(N_c)$, $f^{a_{U(1)}bc}=0$, for all $b,c$) 
and therefore it does not couple directly to gluons; 
however, quarks carry charge under it.
(Its coupling strength has to be readjusted from QCD to
QED strength for it to represent a real photon.)
 
The color algebra can easily be carried out
diagrammatically.\cite{DoubleLine}
Starting with any given Feynman diagram, one interprets it
as just the color factor for the full diagram, and then
makes the two substitutions, \eqns{struct}{colorfierz}, 
which are represented diagrammatically in \fig{FabcdabFigure}.   
In \fig{TreeExampleFigure} we use these steps to
simplify a sample diagram for five-gluon scattering at tree level.
The final line is the diagrammatic representation of a single trace,
$\Tr\bigl(T^{a_1}T^{a_2}T^{a_3}T^{a_4}T^{a_5}\bigr)$, plus all possible
permutations.
Notice that the $-1/N_c$ terms in \eqn{colorfierz} do not
contribute here, because the photon does not couple to gluons.

 
\figmac{3.0}{fabc_dab}{FabcdabFigure}{} 
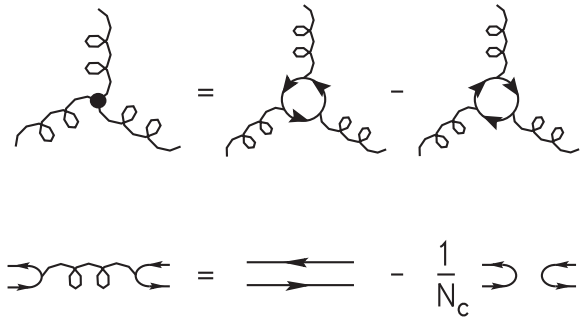
{Diagrammatic equations for simplifying $SU(N_c)$ color algebra.  
Curly lines (``gluon propagators'') represent adjoint indices, 
oriented solid lines (``quark propagators'') represent fundamental 
indices, and ``quark-gluon vertices'' represent the generator matrices 
$(T^a)_i^{~\jb}$.\hfill}


It is easy to see that any tree diagram for $n$-gluon scattering can 
be reduced to a sum of ``single trace'' terms.  This observation leads
to the {\it color decomposition} of the the $n$-gluon
tree amplitude,\cite{TreeColor}
\be \label{eq:treegluecolor}
 {\cal A}^\tree_n \LP \{k_i,\lambda_i,a_i\}\RP
 = g^{n-2} \hskip-1.3mm \sum_{\sigma \in S_n/Z_n} \hskip-1.3mm  
    \Tr\LP T^{a_{\sigma(1)}}\cdots T^{a_{\sigma(n)}}\RP\ 
     A_n^\tree(\sigma(1^{\lambda_1}),\ldots,\sigma(n^{\lambda_n})). 
\ee
Here $g$ is the gauge coupling (${g^2\over4\pi}=\as$), 
$k_i, \lambda_i$ are the gluon momenta and helicities, 
and $A_n^\tree(1^{\lambda_1},\ldots,n^{\lambda_n})$ are the
{\it partial amplitudes}, which contain all the kinematic information.
$S_n$ is the set of all permutations of $n$ objects, while $Z_n$ is
the subset of cyclic permutations, which preserves the trace;
one sums over the set $S_n/Z_n$ in order to sweep out all distinct
cyclic orderings in the trace.
The real work is still to come, in calculating the independent 
partial amplitudes $A_n^\tree$.  However, the partial amplitudes
are simpler than the full amplitude because they are 
{\it color-ordered}:  they only receive contributions from 
diagrams with a particular cyclic ordering of the gluons.
Because of this, the singularities of the partial amplitudes, poles
and (in the loop case) cuts, can only occur in a limited
set of momentum channels, those made out of sums of cyclically adjacent
momenta.  For example, the five-point partial amplitudes
$A_5^\tree(1^{\lambda_1},2^{\lambda_2},3^{\lambda_3},
4^{\lambda_4},5^{\lambda_5})$ can only have poles in 
$s_{12}$, $s_{23}$, $s_{34}$, $s_{45}$, and $s_{51}$,
and not in $s_{13}$, $s_{24}$, $s_{35}$, $s_{41}$, or $s_{52}$,
where $s_{ij} \equiv (k_i+k_j)^2$.

Similarly, tree amplitudes $\qb qgg\cdots g$ with two external
quarks can be reduced to single strings of $T^a$ matrices,
\be \label{eq:treequarkgluecolor}
 {\cal A}^\tree_n 
 = g^{n-2} \hskip-1.3mm \sum_{\sigma \in S_{n-2}} \hskip-1.3mm  
   \LP T^{a_{\sigma(3)}}\cdots T^{a_{\sigma(n)}}\RP_{i_2}^{~\jb_1} 
     A_n^\tree(1_\qb^{\lambda_1},2_q^{\lambda_2},
      \sigma(3^{\lambda_3}),\ldots,\sigma(n^{\lambda_n})), 
\ee
where numbers without subscripts refer to gluons. 
\par\noindent
{\bf Exercise:} Write down the color decomposition for the tree
amplitude $\qb q\Qb Qg$.  

 
\figmac{3.5}{tree_example}{TreeExampleFigure}{} 
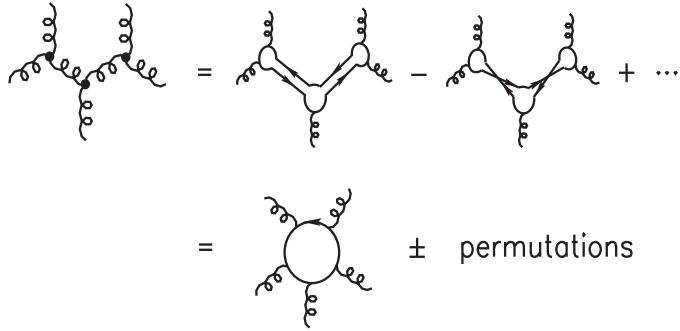
{A sample diagram for tree-level five-gluon scattering, reduced 
 to a single trace.\hfill}


Color decompositions at loop level are equally straightforward.
In \fig{LoopExampleFigure} we simplify a sample diagram for four-gluon 
scattering at one loop.
Again the $-1/N_c$ terms in \eqn{colorfierz} are not
present, but now both single and double trace structures are generated,
leading to the one-loop color decomposition,\cite{LoopColor}  
\bea
&& \hskip-8mm{\cal A}^\oneloop_n \LP \{k_i,\lambda_i,a_i\}\RP 
\nonumber \\
&& \hskip 6mm 
= g^n\Biggl[
    \sum_{\sigma \in S_n/Z_n} 
    N_c\,\Tr\LP T^{a_{\sigma(1)}}\cdots T^{a_{\sigma(n)}}\RP\ 
     A_{n;1}(\sigma(1^{\lambda_1}),\ldots,\sigma(n^{\lambda_n})) 
\nonumber \\
&& \hskip 15mm 
 +\ \sum_{c=2}^{\lfloor{n/2}\rfloor+1}
      \sum_{\sigma \in S_n/S_{n;c}}
    \Tr\LP T^{a_{\sigma(1)}}\cdots T^{a_{\sigma(c-1)}}\RP\ 
    \Tr\LP T^{a_{\sigma(c)}}\cdots T^{a_{\sigma(n)}}\RP\ 
\nonumber \\
\label{eq:loopgluecolor}     
&& \hskip 53mm
 \times\ A_{n;c}(\sigma(1^{\lambda_1}),\ldots,\sigma(n^{\lambda_n}))
     \Biggr]\, , 
\eea      
where $A_{n;c}$ are the partial amplitudes, 
$Z_n$ and $S_{n;c}$ are the subsets of $S_n$
that leave the corresponding single and double trace structures 
invariant, and $\lfloor x \rfloor$ is the greatest integer less than or
equal to $x$.  

 
\figmac{3.5}{loop_example}{LoopExampleFigure}{} 
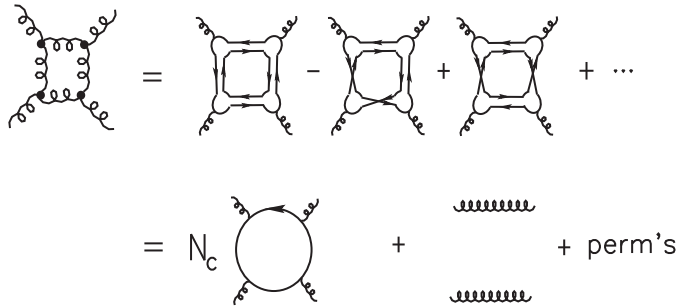
{A diagram for one-loop four-gluon scattering, reduced to single 
 and double traces.\hfill}


The $A_{n;1}$ are the more basic objects in \eqn{loopgluecolor}, 
and are called {\it primitive amplitudes}, because:  
\par\noindent
{\it a}. Like the tree partial amplitudes $A_n^\tree$ in
\eqn{treegluecolor}, they are color-ordered.
\par\noindent
{\it b}. It turns out that the remaining $A_{n;c>1}$ can be 
generated\,\cite{LoopColor,SusyFour} as sums of permutations 
of the $A_{n;1}$.  (For amplitudes with external quarks as well as gluons, 
the primitive amplitudes are not a subset of the partial amplitudes;
new color-ordered objects have to be defined.\cite{qqggg})

One might worry that the color and helicity decompositions
will lead to a huge proliferation in the number of primitive/partial
amplitudes that have to be computed.   Actually it is not too bad,
thanks to symmetries such as parity --- which allows one to
simultaneously reverse all helicities in an amplitude ---
and charge conjugation --- which allows one to exchange a quark and
anti-quark, or equivalently flip the helicity on a quark line.
For example, using parity and cyclic ($Z_5$) symmetry, 
the five-gluon amplitude has only four independent tree-level partial
amplitudes:
\bea  
&& A_5^\tree(1^+,2^+,3^+,4^+,5^+),\qquad\qquad
A_5^\tree(1^-,2^+,3^+,4^+,5^+),
\nonumber \\
\label{eq:treefiveg}
&& A_5^\tree(1^-,2^-,3^+,4^+,5^+),\qquad\qquad 
A_5^\tree(1^-,2^+,3^-,4^+,5^+).
\eea
In fact, we'll see that the first two tree partial amplitudes
vanish, and there is a group theory relation between the last two,
so there is only one independent nonvanishing object to calculate.
At one-loop there are four independent objects --- \eqn{treefiveg} with
$A_5^\tree$ replaced by $A_{5;1}$ --- but only the last two
contribute to the NLO cross-section, due to the tree-level vanishings. 

The group theory relation just mentioned derives from the fact that
the tree color decomposition, \eqn{treegluecolor}, is equally valid
for gauge group $U(N_c)$ as $SU(N_c)$, but any amplitude
containing the extra $U(1)$ photon must vanish.   Hence if we
substitute the $U(1)$ generator --- the identity matrix --- into
the right-hand-side of \eqn{treegluecolor}, and collect 
the terms with the same remaining color structure, that linear
combination of partial amplitudes must vanish.  We get 
\bea
0 &=& A_n^\tree(1,2,3,\ldots,n) + A_n^\tree(2,1,3,\ldots,n) 
 + A_n^\tree(2,3,1,\ldots,n) 
\nonumber \\
\label{eq:treephotondecouple}
&& \hskip 10mm
 + \cdots + A_n^\tree(2,3,\ldots,1,n),   
\eea
often called a ``photon decoupling equation''\cite{LoopColor}
or ``dual Ward identity''\cite{ManganoReview} 
(because \eqn{treephotondecouple} can be derived from string theory,
a.k.a. dual theory).
In the five-point case, we can use \eqn{treephotondecouple} to get
\bea
A_5^\tree(1^-,2^+,3^-,4^+,5^+) &=& - A_5^\tree(1^-,3^-,2^+,4^+,5^+) 
\nonumber \\
&& - A_5^\tree(1^-,3^-,4^+,2^+,5^+)
\nonumber \\
\label{eq:fivephotondecouple}
&& - A_5^\tree(1^-,3^-,4^+,5^+,2^+). 
\eea
The partial amplitude where the two negative helicities
are not adjacent has been expressed in terms of the partial 
amplitude where they are adjacent, as desired.

Since color is confined and unobservable, 
the QCD-improved parton model cross-sections
of interest to us are averaged over initial colors and summed over
final colors.  These color sums can be performed very easily
using the diagrammatic techniques.  For example, 
\fig{ColorSumFigure} illustrates the evaluation of the color sums
needed for the tree-level four-gluon cross-section.
In this case we can use the much simpler $U(N_c)$ color algebra,
omitting the $-1/N_c$ term in~\eqn{colorfierz}, 
because the $U(1)$ contribution vanishes.  (This shortcut is not
valid for general loop amplitudes, or if external quarks are present.)  
Using also the reflection identity discussed below, \eqn{reflectionid}, 
the total color sum becomes
\bea
\sum_{{\rm colors}} [\A_4^{{\rm tree}\,*} \A_4^\tree]
   &=& 2\,g^4\, A_4^{{\rm tree}\,*}(1,2,3,4)
  \times \biggl[ A_4^\tree(1,2,3,4) ( N_c^4 + N_c^2 ) 
\nonumber \\
&&\quad 
   + \Bigl( A_4^\tree(2,1,3,4) + A_4^\tree(2,3,1,4) \Bigr)
          ( N_c^2 + N_c^2 ) \biggr]
\nonumber \\
&&\qquad\qquad +\ \hbox{2 more permutations} 
\nonumber \\ 
&&
\nonumber \\ 
\label{eq:fourgluesum}
&=& g^4 \, N_c^2 (N_c^2-1) \sum_{\si\in S_3} 
  |A_4^\tree(\si(1),\si(2),\si(3),4)|^2\ , 
\eea
where we have used the decoupling identity, \eqn{treephotondecouple},
in the last step.     
 
 
\figmac{2.0}{colorsum}{ColorSumFigure}{} 
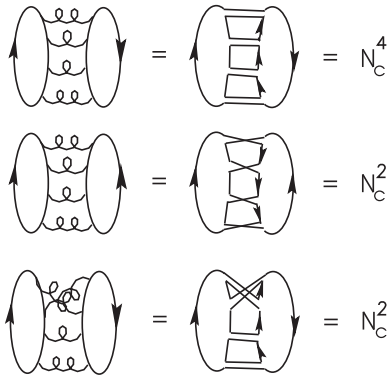
{Diagrammatic evaluation of color sums for the tree-level 
 four-gluon cross-section.\hfill}


Because we have stripped all the color factors out of the partial 
amplitudes, the {\it color-ordered Feynman rules} for constructing 
these objects are purely kinematic (no $T^a$'s or $f^{abc}$'s are
left).   The rules are given in \fig{RulesFigure}, for quantization
in Lorentz-Feynman gauge.  (Later we will discuss alternate gauges.)
To compute a tree partial amplitude, or a {\it color-ordered} loop
partial amplitude such as $A_{n;1}$,
\par\noindent
1. Draw all {\it color-ordered graphs}, i.e. all planar graphs where 
the cyclic ordering of the 
external legs matches the ordering of the $T^{a_i}$ matrices 
in the corresponding color structure,
\par\noindent
2. Evaluate each graph using the color-ordered vertices of
\fig{RulesFigure}.
\par\noindent
Starting with the standard Feynman rules in terms of $f^{abc}$, etc.,
you can check that this prescription works because:
\par\noindent
1) of all possible graphs, only the color-ordered graphs can 
contribute to the desired color structure, and
\par\noindent
2) the color-ordered vertices are obtained by inserting
\eqn{struct} into the standard Feynman rules and extracting a single
ordering of the $T^a$'s; hence they keep only the portion of a
color-ordered graph which does contribute to the correct color 
structure. 
\par\noindent
Many partial amplitudes are {\it not} color-ordered --- for example
the $A_{n;c}$ for $c>1$ in \eqn{loopgluecolor} --- and so the above
rules do not apply.  However, as mentioned above one can usually 
express such quantities as sums over permutations of color-ordered
``primitive amplitudes'' --- for example the $A_{n;1}$ --- to which 
the rules do apply. 

 
\figmac{3.2}{rules}{RulesFigure}{} 
{Color-ordered Feynman rules, in Lorentz-Feynman gauge, omitting
 ghosts. Straight lines represent fermions, wavy lines gluons.  
 All momenta are taken outgoing.\hfill}


\subsection{Helicity Nitty Gritty}

The spinor helicity formalism for massless vector
bosons\,\cite{SpinorHelicity,XuZhangChang,GunionKunszt} is largely 
responsible for the existence of extremely compact representations 
of tree and loop partial amplitudes in QCD.  
It introduces a new set of kinematic objects,
spinor products, which neatly capture the collinear behavior of these
amplitudes.  A (small) price to pay is that automated simplification of
large expressions containing these objects is not always
straightforward, because they obey nonlinear identities.
In this section we will review the spinor helicity formalism and 
some of the key identities.

We begin with massless fermions. 
Positive and negative energy solutions of the massless Dirac
equation are identical up to normalization conventions.  
One way to see this is to note that the positive and 
negative energy projection operators, 
$\Lambda_+(k) \sim u(k)\otimes \overline{u(k)}$ and
$\Lambda_-(k) \sim v(k)\otimes \overline{v(k)}$, are both proportional
to $\ksl$ in the massless limit.
Thus the solutions of definite helicity,
$u_\pm(k) = \hf(1\pm\gamma_5)u(k)$ and 
$v_\mp(k) = \hf(1\pm\gamma_5)v(k)$, 
can be chosen to be equal to each other.  
(For negative energy solutions, the helicity is the negative of 
the chirality or $\gamma_5$ eigenvalue.)
A similar relation holds between the conjugate spinors 
$\overline{u_\pm(k)} = \overline{u(k)}\hf(1\mp\gamma_5)$
and $\overline{v_\mp(k)} = \overline{v(k)}\hf(1\mp\gamma_5)$.
Since we will be interested in amplitudes with a large number of
momenta, we label them by $k_i$, $i=1,2,\ldots,n$, and use the 
shorthand notation
\be \label{eq:spinorshort}
|i^\pm\rangle\ \equiv\ |k_i^\pm\rangle\ \equiv\ 
  u_\pm(k_i)\ =\ v_\mp(k_i), \qquad 
\langle i^\pm| \equiv\ \langle k_i^\pm| \equiv\ 
\overline{u_\pm(k_i)}
   \ =\ \overline{v_\mp(k_i)}. 
\ee
We define the basic spinor products by
\be \label{eq:basicspinordef}
\spa{i}.{j}\ \equiv\ \langle i^-|j^+\rangle
  \ =\ \overline{u_-(k_i)} u_+(k_j),  \qquad  
\spb{i}.{j}\ \equiv\ \langle i^+|j^-\rangle
  \ =\ \overline{u_+(k_i)} u_-(k_j). 
\ee
The helicity projection implies that products like 
$\langle i^+|j^+ \rangle$ vanish.

For numerical evaluation of the spinor products, it is useful to 
have explicit formulae for them, for some representation of the Dirac
$\gamma$ matrices.  In the Dirac representation,
\be \label{eq:gammamatrices}
\gamma^0\ =\ \left(\matrix{{\bf1}&{\bf0}\cr{\bf0}&-{\bf1}\cr}\right)\ ,
\qquad 
\gamma^i\ =\ \left(\matrix{{\bf0}&{\bf\sigma}^i\cr
                           -{\bf\sigma}^i&{\bf0}\cr}\right)\ ,
\qquad 
\gamma_5\ =\ \left(\matrix{{\bf0}&{\bf1}\cr{\bf1}&{\bf0}\cr}\right)\ ,
\ee
the massless spinors can be chosen as follows,
\be \label{eq:explicitspinor}
u_+(k) = v_-(k) = {1\over\sqrt{2}}
  \left[ \matrix{ \sqrt{k^+} \cr 
   \sqrt{k^-} e^{i\varphi_k} \cr
                  \sqrt{k^+} \cr 
   \sqrt{k^-} e^{i\varphi_k} \cr} \right] , \hskip3mm
u_-(k) = v_+(k) = {1\over\sqrt{2}}
  \left[ \matrix{ \sqrt{k^-} e^{-i\varphi_k} \cr 
                                 -\sqrt{k^+} \cr
                - \sqrt{k^-} e^{-i\varphi_k} \cr 
                                  \sqrt{k^+} \cr} \right] , 
\ee
where 
\be \label{eq:phasekdef}
e^{\pm i\varphi_k}\ \equiv\ 
  { k^1 \pm ik^2 \over \sqrt{(k^1)^2+(k^2)^2} }
\ =\  { k^1 \pm ik^2 \over \sqrt{k^+k^-} }\ ,
\qquad k^\pm\ =\ k^0 \pm k^3.  
\ee
\par\noindent
{\bf Exercise:} Show that these solutions satisfy the massless Dirac
equation with the proper chirality.

Plugging Eqs.~\ref{eq:explicitspinor} into the definitions of the spinor
products, \eqn{basicspinordef}, we get explicit formulae for the case
when both energies are positive,
\bea 
\spa{i}.{j} &=& \sqrt{k_i^- k_j^+} e^{i\varphi_{k_i}}
              - \sqrt{k_i^+ k_j^-} e^{i\varphi_{k_j}}
\ =\ \sqrt{|s_{ij}|} e^{i\phi_{ij}}, 
\nonumber \\ 
\spb{i}.{j} &=& -\sqrt{k_i^- k_j^+} e^{-i\varphi_{k_i}}
              + \sqrt{k_i^+ k_j^-} e^{-i\varphi_{k_j}}
\ =\ \sqrt{|s_{ij}|} e^{-i(\phi_{ij}+\pi)}, 
\nonumber \\
\label{eq:posenergyexplicit} 
&&\qquad\qquad k_i^0 >0, \quad k_j^0 > 0, 
\eea
where  $s_{ij}\ =\ (k_i+k_j)^2\ =\ 2 k_i\cdot k_j$, and  
\be \label{eq:phiijdef}
\cos\phi_{ij}\ =\ { k_i^1 k_j^+ - k_j^1 k_i^+ 
             \over \sqrt{|s_{ij}| k_i^+ k_j^+} }
\ , \qquad
\sin\phi_{ij}\ =\ { k_i^2 k_j^+ - k_j^2 k_i^+ 
             \over \sqrt{|s_{ij}| k_i^+ k_j^+} }
\ .
\ee
The spinor products are, up to a phase, square roots of Lorentz
products.  We'll see that the collinear limits of massless gauge
amplitudes have this kind of square-root singularity, 
which explains why spinor products lead to very compact analytic
representations of gauge amplitudes, as well as improved numerical
stability.

We would like the spinor products to have simple properties under
crossing symmetry, i.e. as energies become negative.\cite{GunionKunszt}  
We define the spinor product $\spa{i}.{j}$ by analytic continuation
from the positive energy case, using the same
formula, \eqn{posenergyexplicit}, but with $k_i$ replaced by $-k_i$ if
$k_i^0 < 0$, and similarly for $k_j$; and with an extra multiplicative 
factor of $i$ for each negative energy particle.
We define $\spb{i}.{j}$ through the identity
\be \label{eq:spbphasedef}
\spa{i}.{j}\spb{j}.{i}
 = \langle i^- | j^+ \rangle \langle j^+ | i^- \rangle
 = \tr\bigl( \hf(1-\gamma_5) \ksl_i \ksl_j \bigr)
 = 2 k_i\cdot k_j = s_{ij}.
\ee
We also have the useful identities:
\par\noindent
Gordon identity and projection operator:  
\be \label{eq:spinornorm}
 \langle i^\pm | \gamma^\mu | i^\pm \rangle\ =\ 2 k_i^\mu,\qquad\qquad
  | i^\pm \rangle \langle i^\pm |\ =\ \hf(1\pm\gamma_5)\ksl_i
\ee
antisymmetry:
\be \label{eq:spinorantisym}
\spa{j}.{i} = -\spa{i}.{j}, \qquad
\spb{j}.{i} = -\spb{i}.{j}, \qquad
\spa{i}.{i} = \spb{i}.{i} = 0 \qquad
\ee
Fierz rearrangement: 
\be \label{eq:spinorfierz}
\langle i^+|\gamma^\mu|j^+\rangle \langle k^+|\gamma_\mu|l^+\rangle
\ =\ 2 \, \spb{i}.{k}\spa{l}.{j}
\ee
charge conjugation of current:
\be \label{eq:cccurrent}
\langle i^+|\gamma^\mu|j^+\rangle
\ =\  \langle j^-|\gamma^\mu|i^-\rangle
\ee
Schouten identity:
\be \label{eq:schouten}
\spa{i}.{j} \spa{k}.{l}\ =\ 
  \spa{i}.{k} \spa{j}.{l} + \spa{i}.{l} \spa{k}.{j} .
\ee
In an $n$-point amplitude, momentum conservation, 
$\sum_{i=1}^n k_i^\mu = 0$, provides one more identity,
\be \label{eq:momcons}
\sum_{{{i=1} \atop {i\neq j,k}}}^n \spb{j}.{i}\spa{i}.{k}\ =\ 0.
\ee

The next step is to introduce a spinor representation for the
polarization vector for a massless gauge boson of definite helicity
$\pm1$,
\be \label{eq:poldefn}
\pol^\pm_\mu(k,q)\ =\ 
 \pm { \langle q^\mp | \gamma_\mu | k^\mp \rangle
  \over \sqrt{2} \langle q^\mp | k^\pm \rangle }\ ,
\ee
where $k$ is the vector boson momentum and $q$ is an auxiliary
massless vector, called the {\it reference momentum}, reflecting the
freedom of on-shell gauge tranformations.
We will not motivate~\eqn{poldefn}, but just show that it has the desired
properties.
Since $\ksl | k^\pm\rangle = 0$, $\pol^\pm(k,q)$ is transverse to $k$, 
for any $q$,
\be \label{eq:transverse}
\pol^\pm(k,q) \cdot k\ =\ 0.
\ee
Complex conjugation reverses the helicity,    
\be \label{eq:polcc}
(\pol_\mu^+)^*\ =\ \pol_\mu^-\ .
\ee
The denominator gives $\pol_\mu$ the standard normalization 
(using~\eqn{spinorfierz}),
\bea \label{eq:polnormnonzero}
\pol^+\cdot(\pol^+)^* &=& \pol^+\cdot\pol^-\ =\ 
-{1\over2} { \langle q^-|\gamma_\mu|k^-\rangle
             \langle q^+|\gamma^\mu|k^+\rangle
             \over \spa{q}.{k} \spb{q}.{k} }\ =\ -1, 
\nonumber \\
\label{eq:polnormzero}
\pol^+\cdot(\pol^-)^* &=& \pol^+\cdot\pol^+\ =\ 
{1\over2} { \langle q^-|\gamma_\mu|k^-\rangle
            \langle q^-|\gamma^\mu|k^-\rangle
             \over {\spa{q}.{k}}^2 }\ =\ 0. 
\eea             
States with helicity $\pm1$ are produced by $\pol^\pm$.  The easiest
way to see this is to consider a rotation around the ${\bf k}$
axis, and notice that the $| k^+ \rangle$ in the denominator
of~\eqn{poldefn} picks up the opposite phase from the state
$| k^- \rangle$ in the numerator; i.e. it doubles the phase from
that appropriate for a spinor (helicity $+\hf$) 
to that appropriate for a vector (helicity $+1$). 
Finally, changing the reference momentum $q$ does amount to an on-shell 
gauge transformation, since $\pol_\mu$ shifts by an amount 
proportional to $k_\mu$:
\bea 
\pol_\mu^+(\tq) - \pol_\mu^+(q) 
\hskip-1.2mm &=& \hskip-1.2mm
{ \langle \tq^- | \gamma_\mu | k^- \rangle
  \over \sqrt{2} \spa{\tq}.{k} }
-  { \langle q^- | \gamma_\mu | k^- \rangle
  \over \sqrt{2} \spa{q}.{k} }
 = - { \langle \tq^- | \gamma_\mu \ksl | q^+ \rangle
       + \langle \tq^- | \ksl \gamma_\mu | q^+ \rangle
  \over \sqrt{2} \spa{\tq}.{k} \spa{q}.{k} } 
\nonumber \\
\label{eq:onshellgauge}                        
\hskip-1.2mm &=& \hskip-1.2mm
- { \sqrt{2} \spa{\tq}.{q}
   \over \spa{\tq}.{k} \spa{q}.{k} } \times k_\mu\ .
\eea
{\bf Exercise:} Show that the completeness relation for these
polarization vectors is that of an light-like axial gauge,
\be \label{eq:polcomplete}
  \sum_{\lambda=\pm} 
  \pol^\lambda_\mu(k,q) \, (\pol^\lambda_\nu(k,q))^*
\ =\ -\eta_{\mu\nu} + { k_\mu q_\nu + k_\nu q_\mu \over k\cdot q }\ .
\ee

A separate reference momentum $q_i$ can be chosen for each gluon 
momentum $k_i$ in an amplitude.  Because it is a gauge choice, 
one should be careful not to change the $q_i$ within the calculation 
of a gauge-invariant quantity (such as a partial amplitude).
On the other hand, different choices can be made when calculating 
different gauge-invariant quantities.
A judicious choice of the $q_i$ can simplify a 
calculation substantially, by making many terms and diagrams vanish,
due primarily to the following identities, where 
$\pol_i^\pm(q) \equiv \pol^\pm(k_i,q_i=q)$:
\bea \label{eq:polvanishidone}
 \pol_i^\pm(q)\cdot q &=& 0, 
\\
\label{eq:polvanishidtwo}
 \pol_i^+(q)\cdot \pol_j^+(q) &=& 
 \pol_i^-(q)\cdot \pol_j^-(q)\ =\ 0, 
\\
\label{eq:polvanishidthree}
 \pol_i^+(k_j)\cdot \pol_j^-(q) &=& 
 \pol_i^+(q)\cdot \pol_j^-(k_i)\ =\ 0, 
\\
\label{eq:polvanishidfour}
 \esl_i^+(k_j) | j^+ \rangle &=& 
 \esl_i^-(k_j) | j^- \rangle\ =\ 0, 
\\
\label{eq:polvanishidfive}
\langle j^+ | \esl_i^-(k_j) &=& 
\langle j^- | \esl_i^+(k_j)\ =\ 0. 
\eea
In particular, it is useful to choose the reference momenta of 
like-helicity gluons to be identical, and to equal the external 
momentum of one of the opposite-helicity set of gluons. 
 
We can now express any amplitude with massless external fermions
and vector bosons in terms of spinor products.   
Since these products are defined for both positive- and negative-energy 
four-momenta, we can use crossing symmetry to extract a number of 
scattering amplitudes from the same expression, by exchanging which 
momenta are outgoing and which incoming.
However, because the helicity of a positive-energy (negative-energy)
massless spinor has the same (opposite) sign as its chirality, 
the helicities assigned to the particles --- bosons as well as fermions
--- depend on whether they are incoming or outgoing.   
Our convention is to label particles with their helicity when they
are considered outgoing (positive-energy); if they are incoming the
helicity is reversed.

The spinor-product representation of an amplitude can be related to
a more conventional one in terms of Lorentz-invariant objects,
the momentum invariants $k_i\cdot k_j$ and contractions of the 
Levi-Civita tensor $\pol_{\mu\nu\sigma\rho}$ with external momenta.
The spinor products carry around a number of phases.  Some of 
the phases are unphysical because they are associated with 
external-state conventions, such as the definitions of the spinors 
$|i^\pm\rangle$.  Physical quantities such as cross-sections
(or amplitudes from which an overall phase has been removed),
when constructed out of the spinor products,  
will be independent of such choices.  
Thus for each external momentum label $i$, if the product 
$\spa{i}.{j}$ appears then its phase should be compensated by 
some $\spb{i}.{k}$ 
(or equivalently $1/\spa{i}.{k}=-\spb{i}.{k}/s_{ik}$).  
If a spinor string appears in a physical quantity, 
then it must terminate, i.e. it has the form
\be \label{eq:longspstring}
\spa{i_1}.{i_2}\spb{i_2}.{i_3}\spa{i_3}.{i_4}\cdots \spb{i_{2m}}.{i_1},
\ee
for some $m$.
Multiplying \eqn{longspstring} by
$1=\spb{i_4}.{i_1}\spa{i_1}.{i_4}/s_{i_1i_4}$, etc.,
we can break up any spinor string into strings of length two and four;
the former are just $s_{ij}$'s (\eqn{spbphasedef}), 
while the latter can then be evaluated by performing the Dirac trace:
\bea
\spa{i}.{j}\spb{j}.{l}\spa{l}.{m}\spb{m}.{i} &=&
\tr\Bigl( \hf(1-\gamma_5) \ksl_i \ksl_j \ksl_l \ksl_m \Bigr)
\nonumber \\
\label{eq:fourspstring} 
&=& {1\over2} \biggl[ s_{ij}s_{lm} - s_{il}s_{jm} + s_{im}s_{jl}
  - 4i \pol(i,j,l,m) \biggr] ,
\eea
where $\pol(i,j,l,m)\ =\ 
\pol_{\mu\nu\sigma\rho} k_i^\mu k_j^\nu k_l^\rho k_m^\sigma$.
Thus the Levi-Civita contractions are always accompanied by an $i$ 
and account for the physical phases.
In practice, the spinor products offer the most compact representation
of helicity amplitudes, but it is useful to know the connection 
to a more conventional representation.   
\par\noindent
{\bf Exercise:}  Verify the Schouten identity, \eqn{schouten},
by multiplying both sides by $\spb{j}.{k}\spb{l}.{i}$ and 
using \eqn{fourspstring} to simplify.


\section{Tree-level techniques}

Now we are ready to attack some tree amplitudes, beginning with
direct calculation of some simple examples, followed by a discussion
of recursive techniques for generating more complicated amplitudes,
and of the role of supersymmetry and factorization properties
in tree-level QCD.


\subsection{Simple examples}

Let's first compute the four-gluon tree helicity amplitude 
$A_4^\tree(1^+,2^+,3^+,4^+)$.\footnote{%
Although we will refer to the gluons as all having the same positive
helicity, remember that the helicity of the two incoming gluons 
(whichever two they may be) is actually negative.
Hence this scattering process changes the helicity of the gluons
by the maximum possible, $-2 \to +2$.}
Since all the gluons have the same
helicity, if we choose all the reference momenta to be the same
null-vector $q$ we can make all the $\pol_i^+\cdot\pol_j^+$ 
terms vanish according to~\eqn{polvanishidtwo}.   
We can't choose $q$ to equal one of the
external momenta, because that polarization vector would have a 
singular denominator.  But we could choose for example the 
null-vector 
$q^\mu = -2 s_{23} k_1^\mu + (s_{12}-s_{23})(2k_2^\mu+k_3^\mu)$.
Actually we won't need the explicit expression for $q$ here, because
when we start to evaluate the various diagrams, we find that they
always contain at least one $\pol_i\cdot\pol_j$, and therefore every
diagram in this helicity amplitude vanishes identically!  

This result generalizes easily to more external gluons.   
Each nonabelian vertex can contribute at most one momentum
vector $k_i$ to the numerator algebra of the graph, and there are at
most $n-2$ vertices.   Thus there are at most
$n-2$ momentum vectors available to contract with the $n$ polarization
vectors $\pol_i$ (the amplitude is linear in each $\pol_i$).
This means there must be at least one $\pol_i\cdot\pol_j$ contraction,
and so the tree amplitude must vanish whenever we can arrange that all
the $\pol_i\cdot\pol_j$ vanish.  Obviously this can be arranged for
the $n$-gluon amplitudes with all helicities the same,
$A_n^\tree(1^+,2^+,3^+,\ldots,n^+)$, by again taking all the reference
momenta to be identical.  And it can be arranged for 
$A_n^\tree(1^-,2^+,3^+,\ldots,n^+)$ by the reference momentum choice
$q_2 = q_3 = \cdots = q_n = k_1$, $q_1 = k_n$.
Thus we have already computed a large number of (zero) amplitudes, 
\be \label{eq:treevanishing}
  A_n^\tree(1^\pm,2^+,3^+,\ldots,n^+)\ =\ 0.
\ee
\par\noindent
{\bf Exercise:} Use an analogous argument to show that the following 
$\qb qgg\ldots g$ helicity amplitudes also vanish:
\be \label{eq:moretreevanishing}
  A_n^\tree(1_\qb^-,2_q^+,3^+,4^+,\ldots,n^+)\ =\ 0.
\ee
\par\noindent
We'll see later that an ``effective'' supersymmetry\,\cite{NewSWI}
of tree-level QCD is responsible for all these vanishings.    

Next we turn to the (nonzero) helicity amplitude
$A_4^\tree(1^-,2^-,3^+,4^+)$, choosing the reference momenta
$q_1=q_2=k_4$, $q_3=q_4=k_1$, so that only the contraction
$\pol_2^-\cdot\pol_3^+$ is nonzero.   It is easy to see from the
color-ordered rules in \fig{RulesFigure} that 
only one of the three potential graphs contributes,
the one with a gluon exchange in the $s_{12}$ channel.
We get
\bea
&&\hskip -4mm
A_4^\tree(1^-,2^-,3^+,4^+)
\nonumber \\
&=& 
\left({i\over\sqrt{2}}\right)^2 \left({-i\over s_{12}}\right)
\nonumber \\
&& \hskip 0mm 
\times \left[ \pol_1^-\cdot\pol_2^- (k_1-k_2)^\mu
     + (\pol_2^-)^\mu \pol_1^-\cdot(2k_2+k_1)
     + (\pol_1^-)^\mu \pol_2^-\cdot(-2k_1-k_2) \right] 
\nonumber \\
&& \hskip 0mm 
\times \left[ \pol_3^+\cdot\pol_4^+ (k_3-k_4)_\mu
     + (\pol_4^+)_\mu \pol_3^+\cdot(2k_4+k_3)
     + (\pol_3^+)_\mu \pol_4^+\cdot(-2k_3-k_4) \right] 
\nonumber \\
&=& -{2i\over s_{12}} \bigl( \pol_2^-\cdot\pol_3^+ \bigr)
   \bigl( \pol_1^-\cdot k_2 \bigr)
   \bigl( \pol_4^+\cdot k_3 \bigr)
\nonumber \\
&=& -{2i\over s_{12}}
  \left(-{2\over2}{\spb4.3\spa1.2 \over \spb4.2\spa1.3}\right) 
  \left(-{\spb4.2\spa2.1\over\sqrt{2}\spb4.1}\right)
  \left(+{\spa1.3\spb3.4\over\sqrt{2}\spa1.4}\right)
\nonumber \\
\label{eq:ggggmmppa}
&=& -i \, {\spa1.2{\spb3.4}^2 \over \spb1.2\spa1.4\spb1.4}\ .
\eea
We can pretty up the answer a bit, using
antisymmetry (\eqn{spinorantisym}), momentum 
conservation (\eqn{momcons}), and $s_{34}=s_{12}$,             
\bea
A_4^\tree(1^-,2^-,3^+,4^+)
&=& -i \, {\spa1.2(\spa2.3\spb3.4)(\spb3.4\spa3.4)
          \over \spb1.2\spa2.3\spa3.4\spa1.4\spb1.4} 
\nonumber \\
&=& i \, { \spa1.2(-\spa2.1\spb1.4)(\spb1.2\spa1.2)
          \over \spb1.2\spa2.3\spa3.4\spa4.1\spb1.4} 
\nonumber \\
\label{eq:ggggmmppb}
&=& i \, { {\spa1.2}^3 \over \spa2.3\spa3.4\spa4.1 }\ , 
\eea
or
\be \label{eq:ggggmmpp}
A_4^\tree(1^-,2^-,3^+,4^+)
\ =\ i \, { {\spa1.2}^4 \over \spa1.2\spa2.3\spa3.4\spa4.1 }\ .        
\ee

The remaining four-gluon helicity amplitude can be obtained from the
decoupling identity, \eqn{treephotondecouple}:
\bea
A_4^\tree(1^-,2^+,3^-,4^+) &=& 
 - A_4^\tree(1^-,3^-,2^+,4^+) - A_4^\tree(1^-,3^-,4^+,2^+) 
\nonumber \\
&=& -i \, \left[ { {\spa1.3}^3 \over \spa3.2\spa2.4\spa4.1 }
               + { {\spa1.3}^3 \over \spa3.4\spa4.2\spa2.1 }
          \right] 
\nonumber \\
\label{eq:ggggmpmpa}
&=& i \, { {\spa1.3}^3 (\spa1.2\spa3.4+\spa1.4\spa2.3)
      \over \spa1.2\spa2.3\spa3.4\spa4.1\spa2.4 }\ , 
\eea
or using the Schouten identity, \eqn{schouten},  
\be \label{eq:ggggmpmp}
A_4^\tree(1^-,2^+,3^-,4^+)\ =\ 
 i \, { {\spa1.3}^4\over\spa1.2\spa2.3\spa3.4\spa4.1 }\ .
\ee
There are no other four-gluon amplitudes to compute, because
parity allows one to reverse all helicities simultaneously,
by exchanging $\spa{}.{} \lr \spb{}.{}$ and multiplying by
$-1$ if there are an odd number of gluons.

Note also that the antisymmetry of the color-ordered rules implies 
that the partial amplitudes (even with external quarks) 
obey a reflection identity,
\be \label{eq:reflectionid}
A_n^\tree(1,2,\ldots,n)\ = (-1)^n\ A_n^\tree(n,\ldots,2,1).
\ee

To obtain the unpolarized, color-summed cross-section for four-gluon
scattering, we insert the nonvanishing helicity amplitudes,
\eqns{ggggmmpp}{ggggmpmp}, into \eqn{fourgluesum}, and sum over the
negative helicity gluons $i,j$:
\bea
\sum_{{{\rm colors}\atop{\rm helicities}}} 
  [\A_4^{{\rm tree}\,*} \A_4^\tree]
\hskip-1.5mm &=& \hskip-1.5mm 
g^4 \, N_c^2 (N_c^2-1) \sum_{i>j=1}^4 \sum_{\si\in S_3} 
  { (s_{ij})^4 \over s_{\si(1)\si(2)} s_{\si(2)\si(3)}
                     s_{\si(3)4} s_{4\si(1)} } . 
\nonumber \\
\label{eq:fourglueexplicitsum}
&&
\eea
Of course polarized cross-sections can be constructed just as easily 
from the helicity amplitudes. 

Next we calculate a sample five-parton tree amplitude,
for two quarks and three gluons, 
$A_5^\tree(1_\qb^-,2_q^+,3^-,4^+,5^+)$, where the momenta without
subscripts label the gluons.
We choose the gluon reference momenta as $q_3=k_2$, $q_4=q_5=k_1$,
so we can use the vanishing relations, 
\eqns{polvanishidfour}{polvanishidfive},  
\be \label{eq:qqgggvanish}
\langle2^+|\esl_3^-\ =\ \esl_4^+ |1^+\rangle
\ =\ \esl_5^+ |1^+\rangle\ =\ 0.
\ee
This kills the graphs where gluons 3 and 5 attach directly
to the fermion line, and the graph with a four-gluon vertex, 
leaving only the two graphs shown in \fig{qqgggFigure}.


\figmac{2.5}{qqggggraphs}{qqgggFigure}{} 
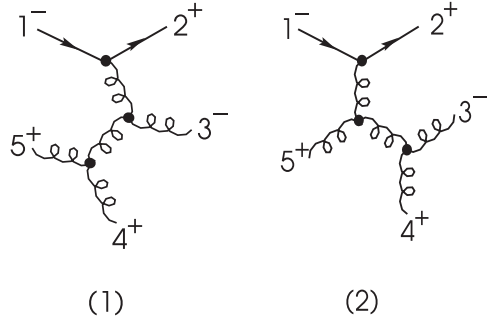
{The two nonvanishing graphs in the $\qb qggg$ helicity amplitude
 calculation.\hfill}


Graph~1 evaluates to 
\bea
&&-{i\over\sqrt{2}} 
{ \langle2^+|(\ksl_3-\ksl_4-\ksl_5)|1^+\rangle \over s_{12} s_{45} }
 (\pol_3^-\cdot \pol_5^+ \pol_4^+\cdot k_5 
- \pol_3^-\cdot \pol_4^+ \pol_5^+\cdot k_4) 
\nonumber \\
&=& -i \, {\spb2.3\spa3.1 \over s_{12}s_{45} }    
 \left[ - {\spb2.5\spa1.3\over\spb2.3\spa1.5}
          {\spa1.5\spb5.4\over\spa1.4}
        + {\spb2.4\spa1.3\over\spb2.3\spa1.4}
          {\spa1.4\spb4.5\over\spa1.5} \right] 
\nonumber \\
&=& +i \, {\spb2.3{\spa1.3}^2\spb4.5 
           \over s_{12}s_{45}\spb2.3\spa1.4\spa1.5 } 
 \Bigl[ - \spa1.5\spb5.2 - \spa1.4\spb4.2 \Bigr] 
\nonumber \\
\label{eq:qqgggone}
&=& -i \, {\spb2.3{\spa1.3}^3\spb4.5 
           \over s_{12}s_{45} \spa1.4\spa1.5 }\ . 
\eea
Graph~2 requires a few more uses of the spinor product identities
({\bf exercise}):
\bea
&& \hskip-8mm
 -{ i\over\sqrt{2} s_{12}s_{34} } \Bigl[
   \langle2^+|(\ksl_3+\ksl_4-\ksl_5)|1^+\rangle 
   \left( \hf \pol_3^-\cdot\pol_4^+ \pol_5^+\cdot(k_3-k_4)
      -\pol_3^-\cdot\pol_5^+ \pol_4^+\cdot k_3 \right) 
\nonumber \\
&&\qquad\qquad\qquad      
 - \langle2^+|(\ksl_3-\ksl_4)|1^+\rangle 
       \pol_3^-\cdot\pol_4^+ \pol_5^+\cdot(k_3+k_4) \Bigr] 
\nonumber \\
\label{eq:qqgggtwo}
&=& \cdots\ =\ +i \, {\spb2.5{\spa1.3}^3\spb3.4 
           \over s_{12}s_{34} \spa1.4\spa1.5 }\ . 
\eea
The sum is 
\be
\label{eq:qqgggsum}
A_5^\tree(1_\qb^-,2_q^+,3^-,4^+,5^+) 
= -i \, { {\spa1.3}^3 \left( - \spb2.3\spa3.4 - \spb2.5\spa5.4 \right)
         \over s_{12}\spa1.4\spa1.5\spa3.4\spa4.5 }\ , 
\ee
or  
\be \label{eq:qqgggmpmpp}
A_5^\tree(1_\qb^-,2_q^+,3^-,4^+,5^+)\ =\ 
  i \, { {\spa1.3}^3 \spa2.3 \over 
    \spa1.2\spa2.3\spa3.4\spa4.5\spa5.1 }\ .
\ee
Once again the expression collapses to a single 
term.\footnote{%
We have multiplied both graphs here by $(-1)$; this external 
state convention makes the $\qb qggg$ partial amplitudes 
equal to the gluino partial amplitudes $\tg\tg ggg$, so that the 
supersymmetry Ward identities below can be applied without 
extra minus signs.}
Spurious singularities associated with the reference momentum choice 
--- such as $1/\spa1.4$ in the above example --- are present in 
individual graphs but cancel out in the gauge-invariant sum.


\subsection{Recursive Techniques}

By now you can see that color-ordering, plus the spinor helicity 
formalism, can vastly reduce the number of diagrams, and terms 
per diagram, that have to be evaluated.  However, with more external 
legs the results still get more complex and difficult to carry out by 
hand.   Fortunately, a technique is available for generating tree 
amplitudes recursively in the number of 
legs.\cite{RecursiveA}  Even if one cannot
simplify analytically the expressions obtained in this way,
the recursive approach lends itself to efficient numerical evaluation.

In order to get a tree-level recursion relation, we need to
construct an auxiliary quantity with one leg off-shell.
For the construction of pure-glue amplitudes,
we define the {\it off-shell current} $J^\mu(1,2,\ldots,n)$
to be the sum of color-ordered ($n+1)$-point Feynman graphs,
where legs $1,2,\ldots,n$ are on-shell gluons, and leg ``$\mu$''
is off-shell, as shown in \fig{GluonCurrentFigure}.
The uncontracted vector index on the off-shell leg is also denoted by
$\mu$; the off-shell propagator is defined to be included in $J^\mu$.
Since $J^\mu$ is an off-shell quantity, it is gauge-dependent.
For example, $J^\mu$ depends on the reference momenta for the 
on-shell gluons, which must therefore be kept fixed until 
after one has extracted an on-shell result.   One can also construct
amplitudes with external quarks recursively, by introducing an off-shell
quark current\,\cite{RecursiveA} as well as the gluon current 
$J^\mu$, but we will not do so here.


\figmac{1.5}{gluoncurrent}{GluonCurrentFigure}{} 
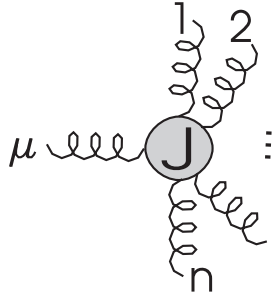
{The off-shell gluon current $J^\mu(1,2,\ldots,n)$.  Leg ``$\mu$'' is
the only off-shell leg.\hfill}


It is easy to write down a recursion relation for $J^\mu$, by
following the off-shell line back into the diagram.
One first encounters either a three-gluon vertex or a four-gluon
vertex.  Each of the off-shell lines branching
out from this vertex attaches to a smaller number of on-shell
gluons, thus we have the recursion relation\cite{RecursiveA}
depicted in \fig{GluonRecursionFigure}, 
\bea
\hskip0mm 
  J^\mu(1,\ldots,n) \hskip-1.5mm &=& \hskip-1.5mm
  {-i\over P^2_{1,n}} \Biggl[ 
  \sum_{i=1}^{n-1} V_3^{\mu\nu\rho}(P_{1,i},P_{i+1,n}) 
  \ J_\nu(1,\ldots,i)\ J_\rho(i+1,\ldots,n)
\nonumber \\
&& \hskip-5mm 
 + \sum_{j=i+1}^{n-1} \sum_{i=1}^{n-2} V_4^{\mu\nu\rho\sigma}
 \ J_\nu(1,\ldots,i)\ J_\rho(i+1,\ldots,j)\ J_\sigma(j+1,\ldots,n)
  \Biggr] ,  
\nonumber \\
\label{eq:Jrecursion}
&& 
\eea
where the $V_i$ are just the color-ordered gluon self-interactions,
\bea 
V_3^{\mu\nu\rho}(P,Q) &=& {i\over\sqrt{2}}
 \left( \eta^{\nu\rho}(P-Q)^\mu + 2\eta^{\rho\mu}Q^\nu 
                            - 2\eta^{\mu\nu}P^\rho \right), 
\nonumber \\
\label{eq:videfn}
V_4^{\mu\nu\rho\sigma} &=& {i\over2} 
 \left( 2\eta^{\mu\rho}\eta^{\nu\sigma} - \eta^{\mu\nu}\eta^{\rho\sigma}
                           - \eta^{\mu\sigma}\eta^{\nu\rho} \right), 
\eea
and
\be \label{eq:pdef}
  P_{i,j}\ \equiv\ k_i+k_{i+1}+\cdots+k_j .
\ee
The $J^\mu$ satisfy the photon decoupling relation,
\be \label{eq:Jdecouple} 
J^\mu(1,2,3,\ldots,n) + J^\mu(2,1,3,\ldots,n) 
 + \cdots + J^\mu(2,3,\ldots,n,1)\ =\ 0,
\ee
the reflection identity
\be \label{eq:Jreflectionid}  
J^\mu(1,2,3,\ldots,n) =\ (-1)^{n+1}\,J^\mu(n,\ldots,3,2,1), 
\ee
and current conservation,
\be \label{eq:Jconserve}
  P^\mu_{1,n}\ J_\mu(1,2,\ldots,n)\ =\ 0.
\ee


\figmac{4.5}{gluonrecursion}{GluonRecursionFigure}{} 
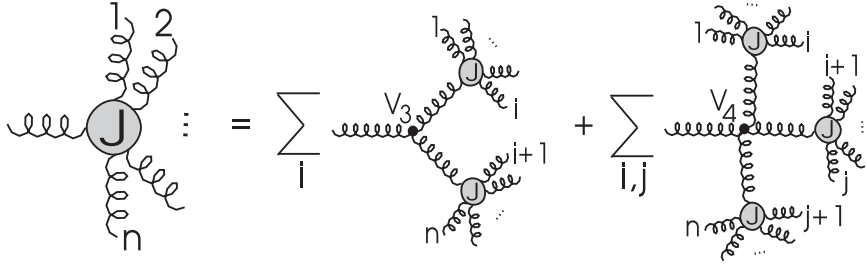
{The recursion relation for the off-shell gluon current 
$J^\mu(1,2,\ldots,n)$.\hfill}


In some cases, the recursion relations can be solved in closed 
form.\cite{RecursiveA,RecursiveB}
The simplest case is (as expected) when all on-shell gluons have the 
same helicity, for which we choose the common reference momentum $q$,
and then
\be \label{eq:Jallplus}
  J^\mu(1^+,2^+,\ldots,n^+)\ =\ 
  { \langle q^-|\gamma^\mu \Psl_{1,n}|q^+\rangle
  \over\sqrt{2} \spa{q}.1\spa1.2\cdots\spa{n-1,}.{n}\spa{n}.{q} }
  \ .
\ee
Let's verify that this expression solves~\eqn{Jrecursion}.
Note first that the $V_4$ term does not contribute at all, nor the
first term in $V_3$, because after
Fierzing we get a factor of $\spa{q}.{q} = 0$.
Thus the right-hand side of~\eqn{Jrecursion} becomes
(using $\spa{q}.{q} = 0$ to commute and rearrange terms)
\bea
&&{ 1 \over \sqrt{2} P^2_{1,n} 
  \spa{q}.1\spa1.2\cdots\spa{n-1,}.{n}\spa{n}.{q} }
 \sum_{i=1}^{n-1} 
   { \spa{i,}.{i+1} \over \spa{i}.{q} \spa{q,}.{i+1} } 
\nonumber \\
&& \times \Bigl( \langle q^-|\gamma^\mu \Psl_{i+1,n}|q^+\rangle 
                \langle q^-|\Psl_{i+1,n}\Psl_{1,i}|q^+\rangle          
\nonumber \\
&&\qquad
             -  \langle q^-|\gamma^\mu \Psl_{1,i}|q^+\rangle 
                \langle q^-|\Psl_{1,i}\Psl_{i+1,n}|q^+\rangle \Bigr)
\nonumber \\
&=&  { \langle q^-|\gamma^\mu \Psl_{1,n}|q^+\rangle
  \over\sqrt{2} P^2_{1,n} 
  \spa{q}.1\spa1.2\cdots\spa{n-1,}.{n}\spa{n}.{q} }
\nonumber \\
\label{eq:checkJrecursion} 
&& \times \left[ \sum_{i=1}^{n-1}
   { \spa{i,}.{i+1} \over \spa{i}.{q} \spa{q,}.{i+1} } 
 \langle q^-| \Psl_{i+1,n} \right] \Psl_{1,n} | q^+\rangle\ .
\eea
Using the identity
\be \label{eq:secondident}
\sum_{i=1}^{n-1}  
   { \spa{i,}.{i+1} \over \spa{i}.{q} \spa{q,}.{i+1} } 
 \langle q^-| \Psl_{i+1,n}\ =\ 
 { \langle 1^-| \Psl_{1,n} \over \spa1.{q} }\ ,
\ee
we get the desired result, \eqn{Jallplus}.
\par\noindent
{\bf Exercise:}  Prove the identity, \eqn{secondident}, by first
proving the identity
\be \label{eq:firstident}
\sum_{i=j}^{k-1}  
   { \spa{i,}.{i+1} \over \spa{i}.{q} \spa{q,}.{i+1} } 
\ =\  { \spa{j}.{k} \over \spa{j}.{q} \spa{q}.{k} }\ .
\ee
The ``eikonal'' identity, \eqn{firstident}, also plays a role in
understanding the structure of the soft singularities of QED
amplitudes, when these are obtained from QCD partial amplitudes by 
the replacement $T^a \to 1$ (see Sections 3.4 and 3.5).

The current where the first on-shell gluon has negative helicity
can be obtained similarly,
\be \label{eq:Joneminus}
  J^\mu(1^-,2^+,\ldots,n^+)\ =\ 
  { \langle 1^-|\gamma^\mu \Psl_{2,n}|1^+\rangle
  \over\sqrt{2} \spa1.2\cdots\spa{n}.{1} }
  \sum_{m=3}^n  { \langle 1^-|\ksl_m \Psl_{1,m}|1^+\rangle
  \over P_{1,m-1}^2 P_{1,m}^2 }\ ,
\ee
where the reference momentum choice is 
$q_1=k_2$, $q_2= \cdots = q_n = k_1$.
\par\noindent
{\bf Exercise:}  Show this.
\par\noindent

Amplitudes with $(n+1)$ legs are obtained from the currents 
$J^\mu(1,2,\ldots,n)$ by amputating the off-shell propagator 
(multiplying by $i \, P^2_{1,n}$), contracting the $\mu$ 
index with the appropriate on-shell polarization vector 
$\pol_{n+1}^\mu$, and taking $P^2_{1,n} = k_{n+1}^2 \to 0$.  
In the case of $J^\mu(1^+,2^+,\ldots,n^+)$, there is no $P^2_{1,n}$ 
pole in the current, so the amplitude must vanish for both
helicities of gluon $(n+1)$, in accord with \eqn{treevanishing}.
In the case of $J^\mu(1^-,2^+,\ldots,n^+)$, the pole term requirement
picks out the term $m=n$ in~\eqn{Joneminus}.  Using reference momentum
$q_{n+1} = k_n$ for $\pol_{n+1}^-$, we obtain 
(replacing $\Psl_{1,n} \to -\ksl_{n+1}$, etc.),
\bea
&& \hskip-15mm A_{n+1}^\tree(1^-,2^+,\ldots,n^+,(n+1)^-) 
\nonumber \\
&=& -i \, 
  { \langle n^+|\gamma_\mu | (n+1)^+\rangle
  \over\sqrt{2}\spb{n,}.{n+1} }  
  { \langle 1^-|\gamma^\mu \Psl_{1,n}|1^+\rangle
  \over\sqrt{2} \spa1.2\cdots\spa{n}.{1} }
   { \langle 1^-|\ksl_n \Psl_{1,n}|1^+\rangle
  \over P^2_{1,n-1} } 
\nonumber \\
\label{eq:mhvinterm}
&=& -i \, { \spa{1,}.{n+1} \over \spa1.2\cdots\spa{n}.{1} }
{ \spa{n+1,}.1\spa1.{n}\spb{n,}.{n+1}\spa{n+1,}.1
  \over s_{n,n+1} }\ , 
\eea
or    
\be \label{eq:mhvadjacent}
  A_n^\tree(1^-,2^-,3^+,4^+,\ldots,n^+)\ =\ 
  i\, { {\spa1.2}^4 \over \spa1.2\cdots\spa{n}.{1} }\ .
\ee
Applying the decoupling identity, \eqn{treephotondecouple}, and the
spinor identity, \eqn{firstident}, it is easy to obtain the remaining
{\it maximally helicity violating} (MHV) or 
Parke-Taylor\,\cite{ParkeTaylor} helicity amplitudes,
\be \label{eq:mhvall}
A_{jk}^\treemhv\ \equiv\ 
A_n^\tree(1^+,\ldots,j^-,\ldots,k^-,\ldots,n^+)\ =\ 
  i\, { {\spa{j}.{k}}^4 \over \spa1.2\cdots\spa{n}.{1} }\ .
\ee

These remarkably simple amplitudes were first conjectured by Parke and
Taylor\,\cite{ParkeTaylor} on the basis of their collinear limits (see
below) and photon decoupling relations, and were rigorously proven
correct by Berends and Giele\,\cite{RecursiveA} using the above 
recursive approach.  The other nonvanishing helicity configurations
(beginning at $n=6$) are typically more complicated.  
The MHV amplitudes can be used as the basis of approximation schemes,
however.\cite{Stirling}

\subsection{Supersymmetry}

What does supersymmetry have to do with a non-supersymmetric theory
such as QCD?  The answer is that tree-level QCD is ``effectively''
supersymmetric,\cite{NewSWI} and the ``non-super\-symmetry'' only leaks 
in at the loop level.   
To see the supersymmetry of an $n$-gluon tree amplitude is
simple:  It has no loops in it, so it has no fermion loops in it.
Therefore the fermions in the theory might as well be gluinos, i.e.
at tree-level the theory might as well be super Yang-Mills theory.
Tree amplitudes with quarks are also supersymmetric, but at the level of
partial amplitudes:  after the color information has been stripped off,
there is nothing to distinguish a quark from a gluino.
Supersymmetry leads to extra relations between
amplitudes, supersymmetric Ward identities (SWI),\cite{OldSWI} 
which can be quite useful in saving computational labor.\cite{NewSWI}

To derive supersymmetric Ward identities,\cite{OldSWI,ManganoReview}
we use the fact that the supercharge $Q$ annihilates the vacuum 
(we are considering exactly supersymmetric theories, 
{\it not} spontaneously or softly broken ones!),
\be \label{eq:startSWI}
0\ =\ \left\langle0\vert [Q, \Phi_1\Phi_2 \cdots \Phi_n] \vert0\right\rangle
\ =\ \sum_{i=1}^n \left\langle0\vert \Phi_1 \cdots [Q,\Phi_i] 
          \cdots \Phi_n \vert0\right\rangle\ .
\ee
When the fields $\Phi_i$ create helicity eigenstates,
many of the $[Q,\Phi_i]$ terms can be arranged to vanish.
To proceed, we need the precise commutation relations of the
supercharge with the fields $g^\pm(k)$, $\Lambda^\pm(k)$, which
create gluon and gluino states of momentum $k$ ($k^2=0$) and helicity
$\pm$.  We multiply $Q$ by a Grassmann spinor parameter $\bar\eta$,
defining $Q(\eta) \equiv \bar\eta^\alpha Q_\alpha$, so that $Q(\eta)$
commutes with the Fermi fields as well as the Bose fields.
The commutators have the form
\bea 
\left[ Q(\eta), g^\pm(k) \right] &=& 
\mp \Gamma^\pm(k,\eta) \, \Lambda^\pm(k), 
\nonumber \\
\label{eq:Qcommutators}   
\left[ Q(\eta), \Lambda^\pm(k) \right] &=& 
\mp \Gamma^\mp(k,\eta) \, g^\pm(k), 
\eea   
where $\Gamma(k,\eta)$ is linear in $\eta$, and has its form
constrained by the Jacobi identity for the supersymmetry algebra,
\be \label{eq:QJacobi}
0\ =\ \left[ \left[ Q(\eta), Q(\zeta) \right], \Phi(k) \right]
 + \left[ \left[ Q(\zeta), \Phi(k) \right],  Q(\eta) \right] 
 + \left[ \left[ \Phi(k), Q(\eta) \right],  Q(\zeta) \right]\ ,
\ee
where $\Phi(k)$ is either $g^\pm(k)$ or $\Lambda^\pm(k)$.
Since $[Q(\eta),Q(\zeta)] = -2i\bar\eta \Psl \zeta$,
we need
\be \label{eq:Gammaconstraint}
  \Gamma^+(k,\eta) \Gamma^-(k,\zeta)  
+ \Gamma^-(k,\eta) \Gamma^+(k,\zeta) 
\ =\ -2i \bar\eta\ksl\zeta.   
\ee

A solution to~\eqn{Gammaconstraint}, which also has the correct
behavior under rotations around the ${\bf k}$ axis, is ({\it cf.}
\eqn{spinornorm})
\be \label{eq:Gammaexpr}
  \Gamma^+(k,\eta)\ =\ \bar\eta u_-(k), \qquad 
  \Gamma^-(k,\eta)\ =\ \bar\eta u_+(k)\ =\ \bar{u}_-(k)\eta.
\ee
Finally, we choose $\eta$ to be a Grassmann parameter $\theta$,
multiplied by the spinor for an arbitrary massless vector $q$, 
and choose $q$ so as to simplify the identities (much 
like the choice of reference momentum in $\pol_\pm^\mu(q)$).
Then $\Gamma^\pm(k,\eta)$ become
\be \label{eq:newGammaexpr}
\Gamma^+(k,q)\ =\ \theta \, \langle q^+ | k^- \rangle 
             \ =\ \theta \, \spb{q}.{k}, \quad
\Gamma^-(k,q)\ =\ \theta \, \langle q^- | k^+ \rangle 
             \ =\ \theta \, \spa{q}.{k}. 
\ee

The simplest case is the like-helicity one.  We start with
\bea 
0 \hskip-2mm &=& \hskip-2mm  
\left\langle0\vert [Q(\eta(q)), \Lambda_1^+ g_2^+ g_3^+ \cdots g_n^+] 
\vert0\right\rangle
\nonumber \\
  \hskip-2mm  &=& \hskip-2mm 
  - \Gamma^-(k_1,q) \, A_n(g_1^+,g_2^+,\ldots,g_n^+)
  + \Gamma^+(k_2,q) \, A_n(\Lambda_1^+,\Lambda_2^+,g_3^+,\ldots,g_n^+)
\nonumber \\
\label{eq:startSWIallplus}
&&\quad 
+\cdots
+\Gamma^+(k_n,q)\,A_n(\Lambda_1^+,g_2^+,\ldots,g_{n-1}^+,\Lambda_n^+).
\eea
Since massless gluinos, like quarks, have only helicity-conserving
interactions in (super) QCD, all of the amplitudes but the first
in~\eqn{startSWIallplus} must vanish.  Therefore so must the 
like-helicity amplitude $A_n(g_1^+,g_2^+,\ldots,g_n^+)$.
Similarly, with one negative helicity we get
\bea 
0 &=& \left\langle0\vert [Q(\eta(q)), \Lambda_1^+ g_2^- g_3^+ \cdots g_n^+] 
      \vert0\right\rangle
\nonumber \\
&=& - \Gamma^-(k_1,q) \, A_n(g_1^+,g_2^-,g_3^+,\ldots,g_n^+)
    - \Gamma^-(k_2,q) \, A_n(\Lambda_1^+,\Lambda_2^-,g_3^+,\ldots,g_n^+),
\nonumber \\
\label{eq:startSWIoneminus}
&&
\eea
where we have omitted the vanishing fermion-helicity-violating 
amplitudes.  Now we use the freedom to choose $q$, setting $q=k_1$
to show the second amplitude vanishes and setting $q=k_2$ 
to show the first vanishes.  Thus we have recovered 
\eqns{treevanishing}{moretreevanishing}.

With two negative helicities, we begin to relate nonzero amplitudes:
\bea
0 &=& 
\left\langle0\vert [Q(\eta(q)), g_1^- g_2^- \Lambda_3^+ 
  g_4^+ \cdots g_n^+] \vert0\right\rangle 
\nonumber \\
 &=& 
  \Gamma^-(k_1,q) \, A_n(\Lambda_1^-,g_2^-,\Lambda_3^+,\ldots,g_n^+)
+ \Gamma^-(k_2,q) \, A_n(g_1^-,\Lambda_2^-,\Lambda_3^+,\ldots,g_n^+)
\nonumber \\
\label{eq:startSWItwominus}
&&\quad - \Gamma^-(k_3,q) \, A_n(g_1^-,g_2^-,g_3^+,\ldots,g_n^+). 
\eea
Choosing $q=k_1$, we get
\be \label{eq:nonzeroSWI}
A_n(g_1^-,g_2^-,g_3^+,g_4^+,\ldots,g_n^+)
\ =\ { \spa1.2\over\spa1.3 } \times 
A_n(g_1^-,\Lambda_2^-,\Lambda_3^+,g_4^+,\ldots,g_n^+).
\ee

No perturbative approximations were made in deriving any of the above 
SWI; thus they hold order-by-order in the loop expansion.
They apply directly to QCD tree amplitudes,
because of their ``effective'' supersymmetry.  
But they can also be used to save some work at the loop level 
(see below).  
Since supersymmetry commutes with color, the SWI apply to each 
color-ordered partial amplitude separately.
Summarizing the above ``MHV'' results (and similar ones including
a pair of external scalar fields), we have
\bea \label{eq:vanishSWI}
 A_n^\susy(1^\pm,2^+,3^+,\ldots,n^+) \hskip-2mm &=& \hskip-2mm 0, 
\\
 A_n^\susy(1^-,2_P^-,3_P^+,4^+,\ldots,n^+)
  \hskip-2mm &=& \hskip-2mm
   \left({\spa1.2\over\spa1.3}\right)^{2|h_P|}
  A_n^\susy(1^-,2_\phi^-,3_\phi^+,4^+,\ldots,n^+). 
\nonumber \\
\label{eq:MHVSWI}
&&
\eea
Here no subscript refers to a gluon, while $\phi$ refers to a scalar 
particle (for which the ``helicity'' $\pm$ means particle vs. antiparticle), 
and $P$ refers to a scalar, fermion or gluon, with respective helicity 
$h_P = 0, \hf, 1$. 

We can use \eqn{MHVSWI} at the four-point level to obtain
the $\qb qgg$ amplitudes from the four-gluon ones, 
\eqns{ggggmmpp}{ggggmpmp}:
\bea \label{eq:qqggmpmp}
  A_4^\tree(1_\qb^-,2_q^+,3^-,4^+) &=& 
    i \,  { {\spa1.3}^3\spa2.3 \over \spa1.2\spa2.3\spa3.4\spa4.1 }\ ,
\nonumber \\
\label{eq:qqggmppm}
  A_4^\tree(1_\qb^-,2_q^+,3^+,4^-) &=& 
    i \,  { {\spa1.4}^3\spa2.4 \over \spa1.2\spa2.3\spa3.4\spa4.1 }\ .
\eea
{\bf Exercise:} Check the SWI at the five-point level, comparing the 
$\qb qggg$ amplitude, \eqn{qqgggmpmpp}, and the $ggggg$ amplitude
from~\eqn{mhvall}.   

\subsection{Factorization Properties}

Analytic properties of amplitudes are very useful as
consistency checks of the correctness of a calculation, but
they can also sometimes be used to help construct amplitudes.
At tree-level, the principal analytic property is the {\it pole}
behavior as kinematic invariants vanish, due to an almost on-shell
intermediate particle.  
As mentioned above, color-ordered amplitudes can only have poles
in channels corresponding to the sum of a sum of {\it cyclically
adjacent} momenta, i.e. as $P^2_{i,j} \to 0$, where
$P^\mu_{i,j} \equiv (k_i+k_{i+1}+\cdots+k_j)^\mu$.
This is because singularities arise from propagators going on-shell,
and propagators for color-ordered graphs always carry momenta of the
form $P^\mu_{i,j}$.
We refer to channels formed by three or more adjacent momenta
as multi-particle channels, and the two-particle channels as
collinear channels.

In a multi-particle channel, a true pole can develop as 
$P_{1,m}^2 \to 0$,
\be \label{eq:multiparticlepole}
  A_n^\tree(1,\ldots,n)\ \sim\ 
\sum_\lambda  A_{m+1}^\tree(1,\ldots,m,P^\lambda) 
        { i \over P^2_{1,m} } 
        A_{n-m+1}^\tree(m+1,\ldots,n,P^{-\lambda}),
\ee
where $P_{1,m}$ is the intermediate momentum and  
$\lambda$ denotes the helicity of the intermediate state $P$.
Our outgoing-particle helicity convention means that the intermediate
helicity is reversed in going from one product amplitude to the other.  

Most multi-parton amplitudes have multi-particle poles, but the MHV 
tree amplitudes do not, due to the vanishing of 
$A_n^\tree(1^\pm,2^+,\ldots,n^+)$.  
When we attempt to factorize an
MHV amplitude on a multi-particle pole, as in \fig{MHVmultipole}(a),
we have only three negative helicities (one from the intermediate
gluon) to distribute among the two product amplitudes.  Therefore one
of the two must vanish, so the pole cannot be present.  Thus the
vanishing SWI also guarantees the simple structure of the nonvanishing
MHV tree amplitudes: only collinear (two-particle) singularities
of adjacent particles are permitted.


\figmac{3.5}{factorization}{MHVmultipole}{} 
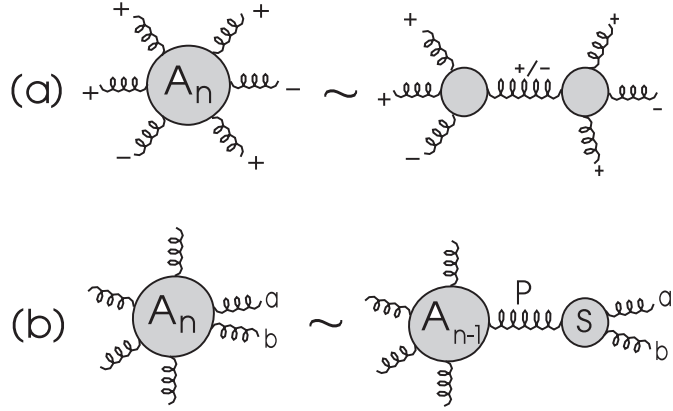
{(a) Factorization of an MHV tree amplitude on a multi-particle pole
--- one of the two product amplitudes always vanishes. 
 (b) General behavior of a tree-level amplitude in the collinear limit
 where $k_a$ is parallel to $k_b$; $S$ stands for the splitting
 amplitude $\Split^\tree$.\hfill}


An angular momentum obstruction suppresses collinear singularities in 
QCD amplitudes.  For example, a helicity $+1$ gluon cannot split into 
two precisely collinear helicity $\pm1$ gluons and still conserve 
angular momentum along the direction of motion.  Nor can it split into 
a $+\hf$ fermion and $-\hf$ antifermion.  The $1/s_{i,i+1}$
from the propagator is cancelled by numerator factors, down to the 
square-root of a pole, 
${1 \over \sqrt{s_{i,i+1}}} \sim {1 \over \spa{i,}.{i+1}} 
\sim {1 \over \spb{i,}.{i+1}}$.  
Thus the spinor products, square roots of Lorentz invariants, are ideal 
for capturing the collinear behavior in QCD.
The general form of the collinear singularities for tree amplitudes is
shown in \fig{MHVmultipole}(b),
\be \label{eq:treesplit}
A_n^\tree(\ldots,a^{\lambda_a},b^{\lambda_b},\ldots)
\ \mathop{\longrightarrow}^{a \parallel b}\
\sum_{\lambda=\pm} 
  \Split^\tree_{-\lambda}(z,a^{\lambda_a},b^{\lambda_b}) \,
  A_{n-1}^\tree(\ldots,P^\lambda,\ldots)\ ,
\ee
where $\Split^\tree$ denotes a {\it splitting amplitude},
the intermediate state $P$ has momentum $k_P=k_a+k_b$
and helicity $\lambda$, and $z$ describes the longitudinal momentum 
sharing, $k_a \approx zk_P$, $k_b \approx (1-z) k_P$.
Universality of the multi-particle and collinear factorization
limits can be derived in field theory,\cite{TreeCollinear} or
perhaps more elegantly in string theory,\cite{ManganoReview} 
which lumps all the field theory diagrams on each side of the 
pole into one string diagram.

An easy way to extract the splitting amplitudes $\Split^\tree$
in \eqn{treesplit} is from the collinear limits of five-point 
amplitudes.  
For example, the limit of $A_5^\tree(1^-,2^-,3^+,4^+,5^+)$ 
as $k_4$ and $k_5$ become parallel determines the gluon splitting 
amplitude $\Split^\tree_-(a^+,b^+)$:
\bea
  A_5^\tree(1^-,2^-,3^+,4^+,5^+) &=& 
   i\, { {\spa1.2}^4 \over \spa1.2\spa2.3\spa3.4\spa4.5\spa5.1 }  
\nonumber \\
  &{ {4 \parallel 5} \atop \mathop{\longrightarrow} } &
  {1 \over \sqrt{z(1-z)} \spa4.5} \times 
  \ i\, { {\spa1.2}^4 \over \spa1.2\spa2.3\spa3.{P}\spa{P}.1 }
\nonumber \\
&=& \Split^\tree_-(4^+,5^+) \times A_4^\tree(1^-,2^-,3^+,P^+).
\nonumber \\
\label{eq:explicitfivetofour} &&
\eea
Using also the $2\parallel3$ and $5\parallel1$ limits, plus parity, 
we can infer the full set of $g\to gg$ splitting 
amplitudes\,\cite{ParkeTaylor,ManganoParke,RecursiveA,ManganoReview}
\bea
\Split^{\rm tree}_{-}(a^{-},b^{-}) &=& 0,
\nonumber \\
\Split^{\rm tree}_{-}(a^{+},b^{+})
            &=& {1\over \sqrt{z (1-z)}\spa{a}.b},
\nonumber \\
\Split^{\rm tree}_{+}(a^{+},b^{-})
            &=& {(1-z)^2\over \sqrt{z (1-z)}\spa{a}.b},
\nonumber \\
\label{eq:gggtree}
\Split^{\rm tree}_{-}(a^{+},b^{-})
            &=& -{z^2\over \sqrt{z (1-z)}\spb{a}.b}.
\eea
The $g\to \qb q$ and $q \to qg$ splitting amplitudes are also easy 
to obtain, from the limits of \eqn{qqgggmpmpp}, etc.

Since the collinear limits of QCD amplitudes are responsible for
parton evolution, it is not surprising that the residue of the 
collinear pole in the square of a splitting amplitude gives the 
(color-stripped) polarized Altarelli-Parisi splitting
probability.\cite{AltarelliParisi}
\par\noindent
{\bf Exercise:} 
Show that the unpolarized $g\to gg$ splitting probability, 
from summing over the terms in \eqn{gggtree}, has the familiar form
\be
 P_{gg}(z)\ \propto\ { 1+z^4+(1-z)^4 \over z(1-z) }\ ,
\ee
neglecting the plus prescription and $\delta(1-z)$ term.  

QCD amplitudes also have universal behavior in the soft limit, where 
all components of a gluon momentum vector $k_s$ go to zero.  
At tree level one finds
\be \label{eq:treesoftlimit}
A_n^\tree(\ldots,a,s,b,\ldots)
\ \mathop{\longrightarrow}^{k_s\to0}\
 \Soft^\tree(a,s,b) \,
  A_{n-1}^\tree(\ldots,a,b,\ldots).
\ee
The soft or ``eikonal'' factor,
\be
 \Soft^\tree(a,s,b)\ =\ { \spa{a}.{b} \over \spa{a}.{s} \spa{s}.{b} }
 \ , 
\ee
depends on both color-ordered neighbors of the soft gluon $s$,
because the sets of graphs where $s$ is radiated from legs 
$a$ and $b$ are both singular in the soft limit.  On the other hand,
the soft behavior is independent of both the identity 
(gluon {\it vs.} quark) and the helicity of partons $a$ and $b$,
reflecting the classical origin of soft radiation.
(See George Sterman's lectures in this volume for a deeper and more
general discussion.\cite{StermanTASI}) 
\par\noindent
{\bf Exercise:} Verify the soft behavior, \eqn{treesoftlimit},
for any of the above multiparton tree amplitudes.

As Zoltan Kunszt will explain in more detail,\cite{KunsztTASI} 
the universal soft and collinear behavior of tree 
amplitudes, and therefore of tree-level cross-sections,
makes possible general procedures for isolating the 
infrared divergences in the real, bremsstrahlung contribution to 
an arbitrary NLO cross-section, and cancelling these divergences
against corresponding ones in one-loop amplitudes.
But the factorization limits also strongly constrain 
the form of tree and loop amplitudes.  
It is quite possible that they uniquely determine a rational
function of the $n$-point variables for $n\geq6$, given the lower-point
amplitudes, but this has not yet been proven.
\par\noindent
{\bf Exercise:} Show that
\be \label{eq:collcounterex}
  { \pol(1,2,3,4) \over \spa1.2\spa2.3\spa3.4\spa4.5\spa5.1 }
\ee
provides a counterexample to the uniqueness assertion at the 
five-point level, because it is nonzero, yet has nonsingular 
collinear limits in all channels.

\subsection{Beyond QCD (briefly)}

This school is titled ``QCD and Beyond'', so let me indicate
briefly how the techniques discussed here can be applied beyond pure 
QCD.  Consider amplitudes containing a single external electroweak
vector boson, $W$, $Z$ or $\gamma$.  
In terms of $U(N_c) = SU(N_c)\times U(1)$ group theory, 
the electroweak boson generator corresponds to the $U(1)$ generator, 
proportional to the identity matrix. 
Thus the color decomposition is identical to that
obtained by ignoring the weak boson.  For example, the tree amplitudes
$\qb qg\cdots g\gamma$ can be written as
\bea 
&& {\cal A}^{\tree,1\gamma}_n \LP 1_\qb,2_q,3,\ldots,n-1,n_\gamma \RP
= \sqrt{2} Q_q e g^{n-3} 
 \hskip-1.3mm \sum_{\sigma \in S_{n-3}} \hskip-1.3mm  
    \LP T^{a_{\sigma(3)}}\cdots T^{a_{\sigma(n-1)}}\RP_{i_1}^{~\ib_2} 
\nonumber \\   
\label{eq:treewithphoton} 
&& \hskip4cm \times  A_n^{\tree,1\gamma}
     (1_\qb,2_q;\sigma(3),\ldots,\sigma(n-1);n_\gamma), 
\eea
where $Q_q$ is the quark charge.  
Furthermore, the partial amplitudes $A_n^{\tree,1\gamma}$ can 
be obtained for free from the partial amplitudes $A_n^\tree$ for 
$\qb qg\cdots g$.   One simply inserts $T^{a_n} = {\bf 1}$ in the 
color decomposition for ${\cal A}_n^\tree$, \eqn{treequarkgluecolor},
and matches the color structures with \eqn{treewithphoton}.   
The result is\,\cite{ManganoReview}
\bea
\hskip-0.5cm
 A_n^{\tree,1\gamma}\LP 1_\qb,2_q;3,\ldots,n-1;n_\gamma \RP
 &=& A_n^\tree\LP 1_\qb,2_q;n,3,4,\ldots,n-1) \RP 
\nonumber \\ 
&& \hskip-0.2cm  + A_n^\tree\LP 1_\qb,2_q;3,n,4,\ldots,n-1) \RP
\nonumber \\ 
   \label{eq:couplingequation}
 + \cdots\hskip-0.7cm && \hskip-0.2cm 
 + A_n^\tree\LP 1_\qb,2_q;3,4,\ldots,n-1,n) \RP .
\eea
Compare this ``photon coupling equation'' with the photon decoupling 
equation for pure gluon amplitudes, \eqn{treephotondecouple}.
When more quark lines are present, one has to pay attention to 
the $-1/N_c$ terms mentioned in Section~2.1, since these distinguish
$SU(N_c)$ from $U(1)$; however, similar formulas can be 
derived, including also multiple photon 
emission.\cite{qqggg,SignerPhoton,SignerThesis}

The emission of a {\it massive} vector particle --- a $W$, $Z$ or 
virtual photon --- would seem to require an extension of the 
helicity formalism of Section~2.2.  However, in most cases one is
actually interested in processes where the vector boson ``decays''
to a pair of massless fermions.  
(One or more of these fermions may be in the
initial state.)  Then the formalism for massless fermions and 
vectors can still be applied, albeit with the introduction of one
additional (but physical) four-vector.   Thus electroweak processes
such as $e^+e^-$ annihilation to four fermions may be calculated
very efficiently using the helicity formalism.

Massive fermions do require a serious extension of the formalism.  
It is possible to represent a massive spinor in terms of
two massless ones\,\cite{MassiveFermionHel}; alternatively one can
represent massive spinor outer products in terms of ``spin 
vectors''.\cite{VegaWudka}  In either case the price is
at least one additional four-vector, this time an unphysical one.
Not only is the formalism more cumbersome than for massless fermions, 
but so are the results.  Amplitudes with a helicity flip on the quark
line no longer vanish; nor do those that were protected by a 
supersymmetry Ward identity in the massless case, such as
$A_4^\tree(1_\qb,2_q,3^+,4^+)$.


\section{Loop-level techniques}

In order to increase the precision of QCD predictions, we need
to go to next-to-leading-order, and in particular, to have efficient
techniques for computing the one-loop amplitudes which now enter.
Here the algebra gets considerably more complicated, even with the use
of color-ordering and the helicity formalism, because there are more
off-shell lines, and more nonabelian vertices. 
Furthermore, one has to evaluate loop integrals with 
loop momenta inserted in the numerator; reducing these integrals
often requires the inversion of matrices which can generate a big mess.
Although the helicity and color tools are still very useful,
we will need additional tools for organizing loop amplitudes 
in order to minimize the growth of expressions in intermediate steps. 

\subsection{Supersymmetry and background-field gauge}

At loop level, QCD ``knows'' it is not supersymmetric.  However, 
one can still rearrange the sum over internal spins propagating 
around the loop, in order to take advantage of supersymmetry.
For example, for an amplitude with all external gluons, and a gluon
circulating around the loop, we can use 
supersymmetry to trade the internal gluon loop for a scalar loop.   
We rewrite the internal gluon loop $g$ (and fermion loop $f$) as a 
supersymmetric contribution plus a complex scalar loop $s$,
\bea
  g &=& (g+4f+3s)\ -\ 4(f+s)\ +\ s\ =\ A^{N=4}\ -\ 4\,A^{N=1}
  \ +\ A^{\rm scalar}, 
\nonumber \\
\label{eq:SusyDecomp}
  f &=& (f+s)\ -\ s\ =\ A^{N=1}\ -\ A^{\rm scalar}. 
\eea
Here $A^{N=4}$ represents the contribution of the $N=4$ super Yang-Mills
multiplet, which contains a gluon $g$, four gluinos $f$, and 
three complex (six real) scalars $s$; while $A^{N=1}$ gives the
contribution of an $N=1$ chiral matter supermultiplet, one fermion plus
one complex scalar.  The advantages of this decomposition are twofold:
\par\noindent
(1) The supersymmetric terms are much simpler than the
nonsupersymmetric ones; not only do they obey SWIs, but we will see
that they have diagram-by-diagram cancellations built into them.
\par\noindent
(2) The scalar loop, while more complicated
than the supersymmetric components, is algebraically simpler
than the gluon loop, because a scalar cannot propagate spin information 
around the loop. 
\par\noindent 
In the context of TQM, this use of supersymmetry could be termed
``internal spin management''.

As an example of how this rearrangement looks, consider the five-gluon 
primitive amplitude $A_{5;1}(1^-,2^-,3^+,4^+,5^+)$, 
whose components according to \eqn{SusyDecomp} are\,\cite{FiveGluon} 
\bea
  A^{N=4} \hskip-2mm &=& \hskip-2mm 
     \cg \, A^\tree \sum_{j=1}^5 \Biggl[ 
     -{1\over\e^2} \LP {\mu^2\over -s_{j,j+1}}\RP^\e
     + \ln\LP{-s_{j,j+1}\over -s_{j+1,j+2}}\RP\,
       \ln\LP{-s_{j+2,j-2}\over -s_{j-2,j-1}}\RP
     + {\pi^2\over6} \Biggr] 
\nonumber \\
  A^{N=1} \hskip-2mm &=& \hskip-2mm
     \cg \, A^\tree \Biggl[ {1\over\e} 
    +{1\over2}\LB\ln\LP{\mu^2\over -s_{23}}\RP
                +\ln\LP{\mu^2\over -s_{51}}\RP\RB + 2 \Biggr] 
\nonumber \\
&&\quad + {i\cg\over2}
   {{\spa1.2}^2 \LP\spa2.3\spb3.4\spa4.1+\spa2.4\spb4.5\spa5.1\RP\over
    \spa2.3\spa3.4\spa4.5\spa5.1}
     {\ln\LP {-s_{23}\over -s_{51}}\RP\over s_{51}-s_{23}} 
\nonumber \\
  A^{\rm scalar} \hskip-2mm &=& \hskip-2mm 
   {1\over3} A^{N=1} + {2\over9} \cg \, A^\tree 
\nonumber \\
&&\hskip-17mm + {i\cg\over3} \Biggl[
   - { \spb3.4\spa4.1\spa2.4\spb4.5
       \LP\spa2.3\spb3.4\spa4.1+\spa2.4\spb4.5\spa5.1\RP
          \over\spa3.4\spa4.5 } 
\nonumber \\
&&\hskip30mm           
     \times { \ln\LP{-s_{23}\over -s_{51}}\RP
         -{1\over2}\LP{s_{23}\over s_{51}}-{s_{51}\over s_{23}}\RP
               \over (s_{51}-s_{23})^3 } 
\nonumber \\
&& \hskip-17mm
   - {\spa3.5{\spb3.5}^3\over\spb1.2\spb2.3\spa3.4\spa4.5\spb5.1}
   + {\spa1.2{\spb3.5}^2\over\spb2.3\spa3.4\spa4.5\spb5.1}
   + {1\over2}{\spa1.2\spb3.4\spa4.1\spa2.4\spb4.5\over
                  s_{23}\spa3.4\spa4.5 s_{51}} \Biggr] 
\label{eq:gggggmmppploop}
\eea
where $A^\tree = A_5^\tree(1^-,2^-,3^+,4^+,5^+)$ is given 
in \eqn{mhvadjacent}, $\mu$ is the renormalization scale, and 
\be \label{eq:cgammadef}
  c_\Gamma\ =\ {\Gamma(1+\e)\Gamma^2(1-\e)
         \over(4\pi)^{2-\e}\Gamma(1-2\e)}\ .
\ee
These amplitudes contain both infrared and ultraviolet divergences,
which have been regulated dimensionally with $D=4-2\e$, 
dropping ${\cal O}(\e)$ corrections.
We see that the three components have quite different analytic
structure, indicating that the rearrangement is a natural one.  
As promised, the $N=4$ supersymmetric component 
is the simplest, followed by the $N=1$ component.  
The non-supersymmetric scalar component is the most complicated, 
yet it is still simpler than the direct gluon calculation,
because it does not mix all three components together.

We can understand why the supersymmetric decomposition works by
quantizing QCD in a special gauge, background-field 
gauge.\cite{Background}  The color-ordered rules in \fig{RulesFigure}
were obtained using the Lorentz gauge condition 
$\del^\mu A_\mu=0$,
where $A_\mu \equiv A_\mu^a T^a$ with $T^a$ in the fundamental
representation.   
After performing the Faddeev-Popov trick to integrate over the 
gauge-fixing condition, 
one obtains the additional term in the Lagrangian 
\be \label{eq:Lorentzgauge}
   -{1\over2\xi}\Tr (\del^\mu A_\mu)^2,
\ee    
where we chose the integration
weight $\xi=1$ (Lorentz-Feynman gauge) in \fig{RulesFigure}.
To quantize in background-field gauge one splits the gauge field
into a classical background field and a fluctuating quantum field,
$A_\mu\ =\ A_\mu^B + A_\mu^Q$, and imposes the gauge condition 
$D_\mu^B A_\mu^Q = 0$, where 
$D_\mu^B = \del_\mu - \textstyle{i\over\sqrt{2}}gA_\mu^B$ 
is the background-field covariant derivative, with $A_\mu^B$ evaluated
in the adjoint representation.
Now the Faddeev-Popov integration (for $\xi=1$) leads to the additional
term, replacing \eqn{Lorentzgauge},
\be \label{eq:Backgroundgauge}
  -{1\over2}\Tr (D_\mu^B A_\mu^Q)^2 
\ =\ -{1\over2}\Tr (\del_\mu A_\mu^Q 
       - \textstyle{i\over\sqrt{2}} g [A_\mu^B,A_\mu^Q])^2.
\ee    

For one-loop calculations we require
only the terms in the Lagrangian that are quadratic in the quantum
field $A_\mu^Q$; $A_\mu^Q$ describes the gluon propagating around the
loop, while $A_\mu^B$ corresponds to the external gluons.  
Expanding out the classical Lagrangian $-{1\over4}\Tr(F_{\mu\nu}^2)$
plus \eqn{Backgroundgauge}, one finds that the three-gluon ($QQB$) and
four-gluon ($QQBB$) color-ordered vertices are modified from those 
shown in \fig{RulesFigure} to 
\bea 
  V^{QQB}_{\mu\nu\rho} &=& {i\over\sqrt{2}}
  \Bigl[ \eta_{\mu\nu}(k-p)_\rho 
     - 2\eta_{\rho\nu}q_\mu + 2\eta_{\rho\mu}q_\nu \Bigr] \nonumber \\
 \label{eq:Backgroundrules} 
  V^{QQBB}_{\mu\nu\rho\lambda} &=& -{i\over2}    
 \Bigl[ \eta_{\mu\nu}\eta_{\rho\lambda}
 + 2\eta_{\mu\lambda}\eta_{\nu\rho} 
 - 2\eta_{\mu\rho}\eta_{\nu\lambda} \Bigr]\ ;
\eea
the remaining rules remain the same.
In background-field gauge the interactions of a scalar and of a ghost
with the background field are identical, and are given by
\bea 
    V^{ssB}_{\rho} &=& {i\over\sqrt{2}} (k-p)_\rho \nonumber \\
 \label{eq:MoreBackgroundrules} 
  V^{ssBB}_{\rho\lambda} &=& -{i\over2} \eta_{\rho\lambda}\ ;
\eea
of course a ghost loop has an additional overall minus sign.

Now let's use \eqns{Backgroundrules}{MoreBackgroundrules} to compare 
the gluon and scalar contributions to an $n$-gluon one-loop amplitude,
focusing on the terms with the most factors of the loop momentum
in the numerator of the Feynman diagrams, because these give 
rise to the greatest algebraic complications in explicit computations
(see the next subsection).
The loop momentum only appears in the tri-linear vertices, and
only in the first term in $V^{QQB}_{\mu\nu\rho}$, because $q$ is an
external momentum.
This term matches $V^{ssB}_{\rho}$ up to the $\eta_{\mu\nu}$ factor.
Thus the leading loop-momentum terms for a gluon loop
(including the ghost contribution) are identical to those for a complex 
scalar loop: $\eta^\mu_\mu - 2 = 2$ in $D=4$.  
In dimensional regularization this result is still true if one uses a 
scheme such as dimensional reduction\,\cite{DimensionalReduction}
or four-dimensional helicity,\cite{Long}
which leaves the number of physical gluon helicities fixed at two.
In fact, as we'll see shortly, the difference between a gluon loop and
a complex scalar loop has {\it two} fewer powers of the loop momentum
in the numerator --- at most $m-2$ powers in a diagram with $m$
propagators in the loop, versus $m$ for the gluon or scalar loop alone.
In summary, a gluon loop is a scalar loop ``plus a little bit more''.  

To treat fermion loops in the same way, it is convenient to use a
``second-order formalism'' where the propagator looks more like that 
of a boson.\cite{Mapping,MorganMapping}  
It is not necessary to generate the full Feynman rules;
it suffices to inspect the effective action $\Gamma(A)$, which generates
the one-particle irreducible (1PI) graphs.  Scattering amplitudes are
obtained by attaching tree diagrams to the external legs of 1PI graphs,
but this process does not involve the loop momentum and is identical  
for all internal particle contributions.
The scalar, fermion and gluon contributions to the effective action
(the latter in background-field gauge and including the ghost loop) 
are 
\bea
\Gamma^{\rm scalar}(A) &=& \ln{\rm det}^{-1}_{[0]}\LP D^2 \RP,
\nonumber \\    
\Gamma^{\rm fermion}(A) &=& 
   {1\over2}\ln{\rm det}^{1/2}_{[1/2]}
      \LP D^2-\textstyle{g\over\sqrt{2}} 
              \hf\sigma^{\mu\nu}F_{\mu\nu} \RP, 
\nonumber \\ 
\label{eq:EffectiveAction}   
\Gamma^{\rm gluon}(A) &=& 
 \ln{\rm det}^{-1/2}_{[1]}\LP D^2-\textstyle{g\over\sqrt{2}}
                                  \Sigma^{\mu\nu}F_{\mu\nu}\RP
 + \ln{\rm det}^{}_{[0]}\LP D^2 \RP,
\eea
where $D$ is the covariant derivative, $F$ is the external field 
strength, $\hf\sigma_{\mu\nu}$ ($\Sigma_{\mu\nu}$) is the
spin-${1\over2}$ (spin-1) Lorentz generator, and 
${\rm det}{}_{[J]}$ is the one-loop determinant for a particle of
spin $J$ in the loop.
The fermionic contribution has been rewritten in second-order form
using
\be \label{eq:secondorderone} 
   \ln{\rm det}^{1/2}_{[1/2]}\LP \Dsl \RP 
 = {1\over2} \ln{\rm det}^{1/2}_{[1/2]}\LP \Dsl^2 \RP
\ee  
and
\be \label{eq:secondordertwo}
   \Dsl^2 
 = \hf \{ \Dsl,\Dsl \} + \hf [ \Dsl,\Dsl ]   
 = D^2 - \textstyle{g\over\sqrt{2}} \hf\sigma^{\mu\nu}F_{\mu\nu} \, .
\ee 

We want to compare the leading behavior of each contribution 
in \eqn{EffectiveAction} for large loop momentum $\ell$.
The leading behavior possible for an $m$-point 1PI graph is $\ell^m$,
as we saw above in the gluon and scalar cases.  The leading term
always comes from the $D^2$ term in \eqn{EffectiveAction}, because 
$F_{\mu\nu}$ contains only the external momenta, not the loop momentum.  
Using $\Tr_{[0]}(1)=1,\ \Tr_{[1/2]}(1)=\Tr_{[1]}=4$, we see that 
the $D^2$ term cancels between the scalar and
fermion loop, and between the fermion and gluon loop;
hence it cancels in any supersymmetric linear combination.
Subleading terms in supersymmetric combinations come from using 
one or more factors of $F$ in generating a graph; each $F$ costs one
power of $\ell$.   
Terms with a lone $F$ cancel, thanks to 
$\Tr\sigma_{\mu\nu} = \Tr\Sigma_{\mu\nu} = 0$, 
so the cancellation for an $m$-point 1PI graph is from $\ell^m$ 
down to $\ell^{m-2}$.
In a gauge other than background-field gauge, the cancellations
involving the gluon loop would no longer happen diagram by
diagram. 
\par\noindent
{\bf Exercise:} By comparing the traces of products of two and three 
$\sigma_{\mu\nu}$'s ($\Sigma_{\mu\nu}$'s), show that for $A^{N=4}$ 
the cancellation is all the way down to $\ell^{m-4}$.
\par\noindent

The loop-momentum cancellations are responsible for the much simpler 
structure of the supersymmetric contributions to 
$A_{5;1}(1^-,2^-,3^+,4^+,5^+)$ in \eqn{gggggmmppploop}, 
and similarly for generic $n$-gluon loop amplitudes.
As we sketch in the next subsection, loop integrals with fewer powers 
of the loop momentum in the numerator can be reduced more simply to 
``scalar'' integrals  --- integrals with no loop momenta in the
numerator.   In the (supersymmetric) case where the $m$-point 1PI
graphs have at most $\ell^{m-2}$ behavior, the set of integrals
obtained is so restricted that such an amplitude
can be reconstructed directly from its absorptive parts\,\cite{SusyOne}
(see Section 4.3).

Similar rearrangements can be carried out for one-loop amplitudes with
external fermions.\cite{SusyOne,qqggg} 
For example, the amplitude with two external quarks
and the rest gluons has many diagrams where a fermion goes part of 
the way around the loop, and a gluon the rest of the way around.
It is easy to see that these graphs have an $\ell^{m-1}$ behavior.  
If one now subtracts
from each graph the same graph where a scalar replaces the gluon in
the loop, then the background-field gauge rules,
\eqns{Backgroundrules}{MoreBackgroundrules}, show that the difference
obeys the ``supersymmetric'' $\ell^{m-2}$ criterion (even though
in this case it is not supersymmetric).  Subtracting and adding back 
this scalar contribution is a rearrangement analogous to the $n$-gluon
supersymmetric rearrangement, and does aid practical
calculations.\cite{qqggg} 

Finally, these rearrangements can be motivated by the 
Neveu-Schwarz-Ramond representation of superstring
theory.\cite{StringBased,Long,Mapping,SusyFour}
This representation is not manifestly space-time supersymmetric, 
but at one loop it corresponds to field theory in 
background-field gauge (for 1PI graphs) 
and to a second-order formalism for fermions.\cite{Mapping}
At tree-level --- and at loop-level for the trees that have to be 
sewn onto 1PI graphs to construct amplitudes --- 
string theory corresponds to the 
nonlinear Gervais-Neveu gauge,\cite{GervaisNeveu,Mapping} 
$\del_\mu A_\mu - {i\over\sqrt{2}}g A_\mu A_\mu = 0$. 
This gauge choice also simplifies the respective 
calculations, though we omit the details here.
String theory may have more to teach us about special gauges
at the multi-loop level.

\subsection{Loop Integral Reduction}

Even if one takes advantage of the various techniques already outlined,
loop calculations with many external legs can still be very complex.
Most of the complication arises at the stage of doing the loop 
integrals.  The general one-loop $m$-point integral in $4-2\e$
dimensions (for vanishing internal particle masses) is
\be \label{eq:MPointLoopIntegral}
I_m\bigl[P(\ell^\mu)\bigr] = 
\int{d^{4-2\e}\ell \over (2\pi)^{4-2\e} } 
 { P(\ell^\mu) \over
\ell^2 (\ell-k_1)^2 (\ell-k_1 - k_2)^2 
\cdots (\ell-k_1-k_2 - \cdots - k_{m-1})^2 }
\ee
where $k_i$, $i=1,\ldots,m$, are the momenta flowing out of the loop at
leg $i$, and $P(\ell^\mu)$ is a polynomial in the loop momentum.
As we'll outline, \eqn{MPointLoopIntegral} can be
reduced recursively to a linear combination of 
{\it scalar} integrals $I_m[1]$, where $m=2,3,4$.
The problem is that for large $m$ the 
reduction coefficients can depend on many kinematic variables, and
are often unwieldy and contain spurious singularities.

Here we illustrate one reduction procedure
that works well for large $m$.\cite{VNV}
If $m\geq5$, then for generic kinematics we have at least four 
independent momenta, say $p_1=k_1$, $p_2=k_1+k_2$, $p_3=k_1+k_2+k_3$, 
$p_4=k_1+k_2+k_3+k_4$.  We can define a set of dual momenta $v_i^\mu$,
\bea 
\hskip-0.5cm
&& v_1^\mu = \pol(\mu,2,3,4), \quad 
v_2^\mu = \pol(1,\mu,3,4), \quad 
v_3^\mu = \pol(1,2,\mu,4), \quad 
v_4^\mu = \pol(1,2,3,\mu),  
\nonumber \\
\label{eq:dualmomenta}
\hskip-0.5cm
&&v_i \cdot p_j = \pol(1,2,3,4)\,\delta_{ij},  
\eea
and expand the loop momentum in terms of them,
\bea 
\ell^\mu &=& {1\over \pol(1,2,3,4)} 
   \sum_{i=1}^4 v_i^\mu\ \ell\cdot p_i
\nonumber \\   
 \label{eq:expandloop}
 &=& {1\over 2\pol(1,2,3,4)} 
   \sum_{i=1}^4 v_i^\mu \bigl[ \ell^2 - (\ell-p_i)^2 + p_i^2 \bigr]\ .
\eea     
The first step can be verified by contracting both sides with
$p_j^\mu$.  In the second step we rewrite $\ell^\mu$ in terms of 
the propagator denominators in \eqn{MPointLoopIntegral}, 
plus a term independent of the loop momentum.  
If we insert \eqn{expandloop} into the 
degree $p$ polynomial $P(\ell^\mu)$ in \eqn{MPointLoopIntegral}, 
the former terms cancel propagator denominators, turning an
$m$-point loop integral into $(m-1)$-point integrals with polynomials
of degree $p-1$, while the latter term remains an $m$-point integral,
also of degree $p-1$.
Iterating this procedure, $m$-point integrals can be reduced
to box integrals ($m=4$) plus scalar $m$-point integrals.
Equation~\ref{eq:expandloop} is only valid for the four-dimensional
components of the loop momentum, so one has to be careful when applying
it to dimensionally-regulated amplitudes.  In practice, when
using the helicity formalism the loop momenta usually end up
contracted with four-dimensional external momenta and polarization 
vectors, in which case $\ell^\mu$ is already projected into 
four-dimensions. 

The strategy of rewriting the loop momentum polynomial $P(\ell^\mu)$ 
(which may be contracted with external momenta) in terms of the 
propagator denominators $\ell^2$, $(\ell-k_1)^2$, etc.
is a very general one.  In special cases --- such as the $N=4$ 
supersymmetric example in Section 4.4 --- 
the form of the contracted $P(\ell^\mu)$ often allows a rapid
reduction without having to invoke the general formalism, and
without undue algebra.  However, in other cases one may not be so
fortunate.

The scalar integrals for $m\geq6$ can be reduced to lower-point scalar
integrals by a similar technique.\cite{Melrose,VNV}
For $m\geq6$ we have a fifth independent vector,
$p_5=k_1+k_2+k_3+k_4+k_5$.  Contracting \eqn{expandloop} with $p_5$, 
we get 
\be \label{eq:expandscalarloop}
\ell \cdot p_5\ =\ {1\over \pol(1,2,3,4)} 
   \sum_{i=1}^4 v_i\cdot p_5\ \ell\cdot p_i ,
\ee   
which can be rewritten as an equality relating a sum of six 
propagator denominators to a term independent of the loop momentum.
Inserting this equality into the scalar integral $I_m[1]$, we get an
expression for $I_m[1]$ as a linear combination of six ``daughter'' 
integrals $I_{m-1}^{(i)}[1]$, where the index $(i)$ indicates which 
of the $m$ propagators has been cancelled. 
A similar formula reduces the scalar pentagon to a sum of five 
boxes.\cite{Melrose,VNV,IntegralsShort,IntegralsLong}
To reduce box integrals with loop momenta in the numerator, 
one may employ either a standard Passarino-Veltman
reduction,\cite{PassarinoVeltman} or one using dual vectors
like that discussed above.\cite{Oldenborgh,SignerThesis}
These approaches share the property of \eqn{expandloop}, that
in each step the degree of the loop-momentum polynomial drops by one.
Thus supersymmetric cancellations of $m$-point 1PI graphs
down to $\ell^{m-2}$ are maintained under integral reduction. 

The final results for an amplitude may therefore be described as a 
linear combination of various bubble, triangle and box scalar integrals.  
The biggest problem is that the reduction coefficients from the above 
procedures contain spurious kinematic singularities, 
which should cancel at the end of the day, 
but which can lead to very large intermediate
expressions if one is not careful.  For example, although the 
Levi-Civita contraction $\pol(1,2,3,4)$ appears in the denominator 
of \eqn{expandloop}, it has an unphysical singularity when the 
four momenta $k_i$ become co-planar, so it should not appear 
in the final result.  Despite this fact, the above approach actually
does a good job of keeping the number of terms small, and the requisite
cancellations of $\pol(1,2,3,4)$ denominator factors are not so hard 
to obtain.

\subsection{Unitarity constraints}

In Section 3.4 we discussed the analytic behavior of tree amplitudes,
namely their pole structure.  At the loop level, amplitudes have cuts
as well as poles.  I won't elaborate on the 
factorization (pole) structure of one-loop amplitudes, but they do
exhibit the same kind of universality as tree amplitudes, 
which leads to strong constraints and consistency
checks on calculations.\cite{AllPlus,SusyFour,BernChalmers}   

Unitarity of the $S$-matrix, $S^\dagger S=1$, implies that the
scattering $T$ matrix, defined by $S=1+iT$, obeys 
$(T-T^\dagger)/i = T^\dagger T$.
One can expand this equation perturbatively in $g$, and recognize 
the matrix sum on the right-hand side as including an integration 
over momenta of intermediate states.
Thus the imaginary or absorptive parts of loop amplitudes ---
which contain the branch-cut information --- can be 
determined from phase-space integrals of products of lower-order 
amplitudes.\cite{Cutting} 
For one-loop multi-parton amplitudes, there are several reasons why 
this calculation of the cuts is much easier than a direct loop
calculation:
\par\noindent
$\bullet$ One can simplify the tree amplitudes {\it before} feeding
them into the cut calculation.
\par\noindent
$\bullet$ The tree amplitudes are usually quite simple, because they
possess ``effective'' supersymmetry, even if the full loop amplitudes
do not.
\par\noindent
$\bullet$ One can further use on-shell conditions for the intermediate
legs in evaluating the cuts.  

The catch is that it is not always possible to reconstruct the 
full loop amplitude from its cuts.  In general there can be an
additive ``polynomial ambiguity'' --- besides the usual
logarithms and dilogarithms of loop amplitudes, one may add 
polynomial terms (actually rational functions) in the kinematic 
variables, which cannot be detected by the cuts.
This ambiguity turns out to be absent in one-loop massless 
supersymmetric amplitudes, due to the loop-momentum 
cancellations discussed in Section 4.1.\cite{SusyFour,SusyOne}
For example, in the five-gluon amplitude, \eqn{gggggmmppploop},
all the polynomial terms in both $A^{N=4}$ and $A^{N=1}$ are 
intimately linked to the logarithms, while in $A^{\rm scalar}$
they are not linked.

The polynomial terms in non-supersymmetric one-loop amplitudes 
cannot generally be reconstructed from unitarity cuts evaluated
in four-dimensions.
It is possible to use dimensional analysis to extract the 
${\cal O}(\e^0)$ polynomial terms if one has evaluated the cuts
to ${\cal O}(\e)$ in dimensional regularization,\cite{EpsCuts}
but this task is significantly harder than evaluation to 
${\cal O}(\e^0)$.
In practice, polynomial ambiguities can often be fixed, recursively in 
the number of external legs, by requiring consistent collinear 
factorization of an amplitude in all 
channels.\cite{AllPlus,BernChalmers}



\figmac{3.8}{samesidecut}{SameSideCutFigure}{} 
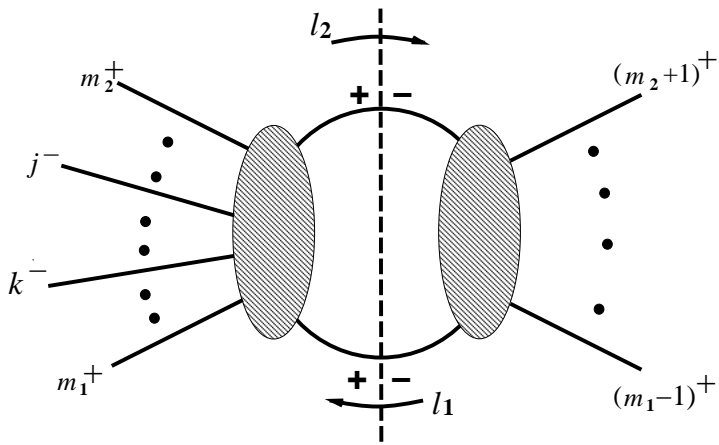
{The possible intermediate helicities for a cut of a MHV amplitude,
when both negative helicity gluons lie on the same side of the cut.
\hfill}


\subsection{Example}

As an example of how simple one-loop multi-parton cuts can be,
we outline here the evaluation of the cuts for an infinite
sequence of $n$-gluon amplitudes, the MHV 
amplitudes in $N=4$ super-Yang-Mills theory.\cite{SusyFour}
We consider the single-trace, leading-color contribution 
$A_{n;1}$, and the case where the two negative helicity gluons lie
on the same side of the cut, as shown in \fig{SameSideCutFigure}.  
(The case where they lie on the opposite side of the cut 
can be quickly reduced to this case\,\cite{SusyFour}
using the SWI, \eqns{vanishSWI}{MHVSWI}.)   
Contributions to this cut from intermediate fermions or scalars 
vanish using the ``effective'' supersymmetry of tree amplitudes,
\eqn{vanishSWI}, plus conservation of fermion helicity and 
scalar particle number, on the right-hand side of the cut.
The only contribution is from intermediate gluons
with the helicity assignment shown in \fig{SameSideCutFigure}.
The tree amplitudes on either side of the cut are
pure-glue MHV tree amplitudes, so using \eqn{mhvall} 
the cut takes the simple form
\bea
\hskip-10mm
  &&\int \dlips(-\ell_1,\ell_2)
   A^\treemhv_{jk}(-\ell_1,m_1,\ldots,m_2,\ell_2) \hfill
\nonumber \\   
\hskip-10mm
&& \hskip3cm \times
  A^\treemhv_{(-\ell_2)\ell_1}(-\ell_2,m_2+1,\ldots,m_2-1,\ell_1) 
\nonumber \\
\hskip-10mm
  &=&  i\, A_{jk}^\treemhv(1,2,\ldots,n)
\nonumber \\
\label{eq:caseacut}
\hskip-10mm
&& \times
\int \dlips(-\ell_1,\ell_2)
    { \spa{m_1-1,}.{m_1} \spa{\ell_1}.{\ell_2}
     \over \spa{m_1-1,}.{\ell_1} \spa{\ell_1}.{m_1} } 
\cdot { \spa{m_2,}.{m_2+1} \spa{\ell_2}.{\ell_1} 
    \over  \spa{m_2}.{\ell_2} \spa{\ell_2,}.{m_2+1} }, 
\eea
where the spinor products are labelled by either loop momenta
($\ell_1,\ell_2$) or external particle labels, and the 
Lorentz-invariant phase space measure for the two-particle intermediate
state is denoted by $\dlips(-\ell_1,\ell_2)$.

The integral in \eqn{caseacut} can be viewed
as a cut hexagon loop integral.
(The four- and five-point cases are degenerate, since there are 
not enough external momenta to make a genuine hexagon.)
To see this, use the on-shell condition $\ell_1^2=\ell_2^2=0$
to rewrite the four spinor product denominators in \eqn{caseacut}
as propagators multiplied by some numerator factor, for
example
\be \label{eq:propagators}
     {1 \over \spa{\ell_1}.{m_1}}
  = {\spb {m_1}.{\ell_1} \over \spa{\ell_1}.{m_1} \spb {m_1}.{\ell_1}}
  = {\spb {m_1}.{\ell_1} \over 2 \ell_1 \cdot k_{m_1}}
  = {-\spb {m_1}.{\ell_1} \over (\ell_1 - k_{m_1})^2}\ .
\ee
In addition to these four propagators, there are two cut propagators
implicit in $\int \dlips(-\ell_1,\ell_2)$.

Rather than evaluate the cut hexagon integral directly,
we use the Schouten identity, \eqn{schouten}, to reduce the number 
of spinor product factors in the
denominator of each term, which will break up the integral
into a sum of cut box integrals.  We have
\be \label{eq:useschoutena}
{ \spa{m_1-1,}.{m_1} \spa{\ell_1}.{\ell_2}
     \over \spa{m_1-1,}.{\ell_1} \spa{\ell_1}.{m_1} } 
 =  { \spa{m_1-1,}.{\ell_2} \over \spa{m_1-1,}.{\ell_1} }
  - { \spa{m_1}.{\ell_2} \over \spa{m_1}.{\ell_1} }\ ,    
\ee
and similarly for the second factor in \eqn{caseacut}.
Four terms are generated, one of which is
\bea     
\hskip-5mm && \hskip-1mm -i \, A_{jk}^\treemhv(1,2,\ldots,n)
\int \dlips(-\ell_1,\ell_2)
    { \spa{m_1}.{\ell_2} \spb{\ell_2}.{m_2}
      \spa{m_2}.{\ell_1} \spb{\ell_1}.{m_1}
\over \spa{m_1}.{\ell_1} \spb{\ell_1}.{m_1}
      \spa{m_2}.{\ell_2} \spb{\ell_2}.{m_2} } \hfill
\nonumber \\
\hskip-5mm
&=& \hskip-1mm -i \, A_{jk}^\treemhv(1,2,\ldots,n)
\int \dlips(-\ell_1,\ell_2)
 { \Tr\Bigl(\hf(1+\gamma_5)\lsl_1 \ksl_{m_1} \lsl_2 \ksl_{m_2}\Bigr)
  \over (\ell_1 - k_{m_1})^2 (\ell_2 + k_{m_2})^2 }\ .
\nonumber \\
\label{eq:firstpartialfr}
&&
\eea
This is the cut box integral $I_4^{m_1,m_2}$, where the set of momenta 
flowing out of its four vertices is
$\{ k_{m_1}, P_{m_1+1,m_2-1}, k_{m_2}, P_{m_2+1,m_1-1} \}$. 
The other three terms similarly give 
$I_4^{m_1-1,m_2}$, $I_4^{m_1,m_2+1}$ and $I_4^{m_1-1,m_2+1}$, 
all with two loop momenta inserted in the numerator.

The $\gamma_5$-odd part of the trace in \eqn{firstpartialfr}
does not contribute, because 
the box does not have enough independent momenta to satisfy the 
Levi-Civita tensor.  The $\gamma_5$-even part can be reduced by
standard Passarino-Veltman techniques\,\cite{PassarinoVeltman} to
scalar box, triangle and bubble integrals.  The coefficient of the 
scalar box integral $I_4^{m_1,m_2}[1]$ is
\be \label{eq:boxcoeff}
 -\hf \bigl( P^2_{m_1,m_2-1} P^2_{m_1+1,m_2} 
          - P^2_{m_1,m_2} P^2_{m_1+1,m_2-1} \bigr)\,.
\ee                
After summing over the four box integrals, the triangles and 
bubbles cancel out.  (This could have been anticipated from the 
exercise in Section 4.1, showing 
that $A^{N=4}$ exhibits loop-momentum cancellations down 
to $\ell^{m-4}$, plus the general loop integral reduction procedures 
discussed in Section 4.2.)
Therefore the $N=4$ MHV amplitude which matches all the cuts 
is a sum of scalar box integrals, with coefficients given by
\eqn{boxcoeff}, which evaluates explicitly (through ${\cal O}(\e^0)$)
to
\be \label{eq:neqfourmhv} 
\hskip-4mm
  A_{n;1}^{N=4}(1^+,\ldots,j^-,\ldots,k^-,\ldots,n^+) = 
  (\mu^2)^\e \, c_\Gamma \, A_{jk}^\treemhv(1,2,\ldots,n) \, V_n\,,
\ee   
where the universal, cyclically symmetric function $V_n$ is given by
\bea
  V_{2m+1} &=& \sum_{r=1}^{m-1} \sum_{i=1}^n f_{i,r}\ ,
\nonumber \\
\label{eq:vneqfour}
  V_{2m} &=& \sum_{r=1}^{m-2} \sum_{i=1}^n f_{i,r}
                            + \sum_{i=1}^{n/2} f_{i,m-1}\ ,
\eea
with
\bea
\hskip-1mm
f_{i,r} \hskip-2mm &=& \hskip-2mm
-{1\over\e^2} \biggl[ \bigl(-P^2_{i-1,i+r-1}\bigr)^{-\e}  
                    + \bigl(-P^2_{i,i+r}\bigr)^{-\e}  
                    - \bigl(-P^2_{i,i+r-1}\bigr)^{-\e}  
                    - \bigl(-P^2_{i-1,i+r}\bigr)^{-\e} \biggr]
\nonumber \\
&& + \Li_2\biggl(1-{P^2_{i,i+r-1} \over P^2_{i-1,i+r-1}}\biggr)                      
   + \Li_2\biggl(1-{P^2_{i,i+r-1} \over P^2_{i,i+r}}\biggr)                      
\nonumber \\
&& + \Li_2\biggl(1-{P^2_{i-1,i+r} \over P^2_{i-1,i+r-1}}\biggr)                      
   + \Li_2\biggl(1-{P^2_{i-1,i+r} \over P^2_{i,i+r}}\biggr)
\nonumber \\
\label{eq:definef}                         
&& - \Li_2\biggl(1 - { P^2_{i,i+r-1} P^2_{i-1,i+r}
                 \over P^2_{i-1,i+r-1} P^2_{i,i+r} }\biggr)  
   + {1\over2} \ln^2
      \biggl( { P^2_{i-1,i+r-1} \over P^2_{i,i+r} } \biggr)\ .
\eea
The dilogarithm is defined by 
$\Li_2(x) = -\int_0^x dt\ \ln(1-t)/t$, and by convention
$(-P^2_{i-1,i-1})^{-\e} = 0^{-\e} = 0$.      


\figmac{3.8}{knowngluonamps}{KnownGluonAmps}{} 
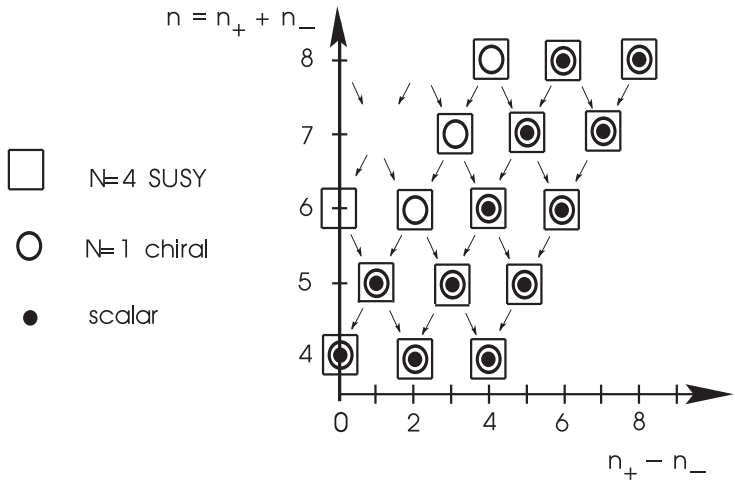
{Currently known one-loop $n$-gluon amplitudes,
decomposed into $N=4$ supersymmetric, $N=1$ chiral, and scalar 
contributions, as in \eqn{SusyDecomp}.
The number of external gluons with helicity $\pm1$ in the amplitude is 
denoted by $n_\pm$.  
Parity reflects the figure about the vertical axis.
Arrows show how amplitudes flow into each other under collinear 
limits.\hfill}


It is remarkable that a compact expression for an infinite 
sequence of gauge theory loop amplitudes is so easy to obtain.
Several other infinite sequences of $n$-gluon one-loop amplitudes
have now been computed, using unitarity as well as collinear and
recursive techniques.\cite{AllPlus,Mahlon,SusyFour,SusyOne}
The currently known $n$-gluon amplitudes --- or rather 
their components under the supersymmetric decomposition
discussed in Section 4.1 ---
are plotted in \fig{KnownGluonAmps} versus the number of helicity
$\pm1$ external states $n_\pm$.
As the figure shows, the supersymmetric components are better known
than the non-supersymmetric scalar terms.  Polynomial ambiguities 
in the non-supersymmetric components of one-loop QCD amplitudes are 
the main obstacle to their efficient evaluation. 
In the various collinear limits, helicity amplitudes (including
their polynomial terms) flow along the arrows in the figure, 
indicating how the limits may be used to help fix the ambiguities.


\section{Conclusions}

In these lectures we described techniques for 
efficient analytical calculation of scattering amplitudes 
in gauge theories, particularly QCD.
Tools such as helicity and color decompositions, special gauges,
unitarity, factorization limits and supersymmetric rearrangements 
can lead to many simplifications.
Some of these ideas can be motivated from string theory, but none
requires its detailed knowledge. 
There is no one ``magic bullet'' but rather a combined arsenal
of techniques that work well together.
At the practical level, some of these tools
have been instrumental in calculating the one-loop five-parton
amplitudes ($ggggg$, $\qb q\qb qg$ and $\qb qggg$) which form the
analytical bottleneck to NLO cross-sections for three-jet events at
hadron colliders.\cite{FiveGluon,Kunsztqqqqg,qqggg} 
They have also been used to obtain infinite sequences of special
one-loop helicity amplitudes in closed
form.\cite{AllPlus,Mahlon,SusyFour,SusyOne}
On the other hand, many processes of experimental interest remain
uncalculated at NLO and at higher orders, so there is plenty of 
room for improvement in the field!


\section*{Acknowledgements}
I would like to thank my collaborators, Zvi Bern, Dave Dunbar and
David Kosower, for contributing greatly to my understanding of
the lecture topics; Zvi Bern, Michael Peskin and Wing Kai Wong for
reading the manuscript; the students at TASI95 for very enjoyable
discussions; and particularly Davison Soper for organizing such a
well-run school. 
This work was supported in part by a NATO Collaborative Research Grant
CRG--921322.
 

\end{document}